\documentclass{aa}  

\usepackage{graphicx}
\usepackage{adjustbox}
\usepackage{txfonts}
\usepackage[]{hyperref}
\begin{document}

   \title{A new sample of super-slowly rotating Ap (ssrAp) stars from the Zwicky Transient Facility survey}

   \author{S.~H{\"u}mmerich\inst{1,2}
          \and
          K.~Bernhard\inst{1,2}
          \and
          E.~Paunzen\inst{3}
          }

   \institute{Bundesdeutsche Arbeitsgemeinschaft f{\"u}r Ver{\"a}nderliche Sterne e.V. (BAV), Munsterdamm 90, 12169 Berlin, Germany
          \and
    American Association of Variable Star Observers (AAVSO), 49 Bay State Rd, Cambridge, MA 02138, USA 
          \and
    Faculty of Science, Masaryk University, Department of Theoretical Physics and Astrophysics, Kotl\'{a}\v{r}sk\'{a} 2, 611\,37 Brno,  Czechia
             }

   \date{Received 2024; accepted 2024}

 
  \abstract
   {The magnetic chemically peculiar Ap stars exhibit an extreme spread of rotational velocities, the reason of which is not well understood. Ap stars with rotational periods of 50 days or longer are know as super-slowly rotating Ap (ssrAp) stars. Photometrically variable Ap stars are commonly termed $\alpha^{2}$ Canum Venaticorum (ACV) variables.}
   {Our study aims at enlarging the sample of known ssrAp stars using data from the Zwicky Transient Facility (ZTF) survey to enable more robust and significant statistical studies of these objects.}
   {Using selection criteria based on the known characteristics of ACV variables, candidate stars were gleaned from the ZTF catalogues of periodic and suspected variable stars and from ZTF raw data. ssrAp stars were identified from this list via their characteristic photometric properties, $\Delta$a photometry, and spectral classification.}
   {The final sample consists of 70 new ssrAp stars, which mostly exhibit rotational periods between 50 and 200 days. The object with the longest period has a rotational period of 2551.7 days. We present astrophysical parameters and a Hertzsprung-Russell diagram for the complete sample of known ssrAp stars. With very few exceptions, the ssrAp stars are grouped in the middle of the main sequence with ages in excess of 150 Myr. ZTF J021309.72+582827.7 was identified as a possible binary star harbouring an Ap star and a cool component, possibly shrouded in dust.}
   {With our study, we enlarge the sample of known ssrAp stars by about 150\%, paving the way for more in-depth statistical studies.}


   \maketitle
%

\section{Introduction}

The Ap/CP2 stars are chemically peculiar stars of the upper main-sequence (spectral types late-B to early-F) defined by the presence of peculiarly strong (or weak) absorption lines of certain elements such as Si, Fe, Cr, Sr, or Eu in their spectra \citep[\textit{e.g.}][]{preston74,RM09}.

They possess strong, globally organised magnetic fields ($\sim$300 G to several tens of kiloGauss; \textit{e.g.} \citealt{babcock58,borra80,mathys91,mathys97,romanyuk14,romanyuk15,bagnulo15,romanyuk16,romanyuk17,romanyuk18,romanyuk20,romanyuk22a,romanyuk22b,romanyuk23}) and exhibit a non-uniform surface distribution of chemical elements, which leads to strictly periodic changes in the
spectra and brightness of many Ap stars that are
satisfactorily explained by the oblique rotator model
\citep{stibbs50}. The observed photometric variability results from a flux redistribution in the unevenly distributed surface abundance patches (``chemical spots'') of certain elements \citep[\textit{e.g.}][]{wolff71,molnar73,krticka12}. This leads to certain peculiarities, such as anti-phase variations between different wavelength regions \citep[\textit{e.g.}][]{molnar73,molnar75,mikulasek07,groebel17} that can be utilised to identify this group of CP stars in the absence of spectroscopic data \citep{faltova21,bf24}.

Photometrically variable Ap stars are commonly termed $\alpha^{2}$ Canum Venaticorum (ACV) variables \citep{morgan33,GCVS1958,IAU,GCVS}. The number of ACV variables with an accurate rotational period determination has increased significantly in recent years through the exploitation of various time-domain photometric surveys. This includes both ground-based surveys \citep[\textit{e.g.}][]{bernhard15,bernhard15b,huemmerich16,bernhard20,faltova21} and satellite missions \citep[\textit{e.g.}][]{paunzen98,huemmerich18,sikora19,paunzen21,labadie-bartz23}.

In comparison to chemically normal main-sequence stars of the same temperature, Ap stars are generally slow rotators. They are, in fact, quite unique in that respect because their rotation periods span five to six orders of magnitude \citep{mathys20a}. While the majority shows periods between 2 and 10 days \citep[\textit{e.g.}][]{netopil17}, some Ap stars have rotation periods as short as $\sim$0.5 days (the current ``record holder'' is HD 60431 with a rotational period of 0.4755 days; \citealt{mikulasek22}). At the other end of the distribution, there is a tail of very slowly rotating Ap stars, with rotation periods of years, decades or even centuries \citep{netopil17,mathys19b,mathys20a,mathys20c,mathys20b,mathys22,mathys24}.

It is currently not well understood why stars of approximately the same evolutionary state show such an extreme spread of rotational velocities. During their main-sequence lifetime, evolutionary changes to the rotation periods of Ap stars are marginal and derive mainly from the conservation of angular momentum \citep[\textit{e.g.}][]{kochukhov06,hubrig07}. Thus, the differentiation in the observed periods is thought to have occurred at the pre-main-sequence stage \citep{mathys22}. Studies of the rotation period distribution of Ap stars, in particular in regard to the tail of very slow rotators, are expected to contribute to the understanding of the origin and evolution of rotational properties in this group of stars as well as provide important input for theoretical studies.

Following \citet{mathys20a}, we refer to Ap stars with rotational periods of 50 d or longer as ``super-slowly rotating Ap'' (ssrAp) stars. Our study aims at enlarging the sample of known ssrAp stars using data from the Zwicky Transient Facility (ZTF) survey \citep{bellm19} to enable more robust and significant statistical studies. Where appropriate, we compare the properties of the new ssrAp stars to the known ones listed in \citet[][Table 1]{mathys24}, who presented a census of the presently known ssrAp stars.

The employed data sources are described in Section \ref{sect:data_sources}. The methods used to identify, classify and investigate our sample stars are detailed in Section \ref{sect:methods}. We present out results in Section \ref{sect:results} and conclude in Section \ref{sect:conclusion}.

\section{Data sources}
\label{sect:data_sources}

This section provides a short overview of the employed data sources.

\subsection{The Zwicky Transient Facility survey} \label{sect:ZTF}

The ZTF is a photometric time-domain survey based at the Palomar Observatory that is regarded as successor to the highly successful Palomar Transient Factory. It has been in operation since 2017 and conducts scans covering 3750 square degrees of sky per hour in three distinct passbands ($g$, $r$, and $i$) to detect objects down to a limiting magnitude of 20.5 mag. To this end, the ZTF camera uses e2v CCD231-C6 devices and is installed on the Palomar 48-inch Samuel Oschin Schmidt Telescope.

The ZTF primary objective is the identification of young supernovae and various other types of transient celestial phenomena. Under its current approach, it collects nearly 300 observations annually for each object. Consequently, ZTF data are also exceptionally well-suited for investigations of variable stars, binary systems, active galactic nuclei, and asteroids. More information on the ZTF survey can be gleaned from \citet{bellm19} and \citet{masci19}.

\subsection{The Large Sky Area Multi-Object Fiber Spectroscopic Telescope (LAMOST)} \label{sect:LAMOST}

The Large Sky Area Multi-Object Fiber Spectroscopic Telescope (LAMOST), also referred to as the Guo Shou Jing telescope \citep{lamost1,lamost2}, is a reflecting Schmidt telescope based at Xinglong Observatory in Beijing, China. It boasts an effective aperture of 3.6$-$4.9\,m and commands a field of view of about 5$\degr$. Due to its unique design, LAMOST is capable of collecting 4000 spectra in a single exposure down to a limiting magnitude of $r$\,$\sim$\,19\,mag and is dedicated to a spectral survey of the entire available sky (about $-10^\circ\,<\,\delta\,<\,+90^\circ$).

We here employ LAMOST low-resolution spectra, which have a spectral resolution of R\,$\sim$\,1800 and cover the wavelength range from 3700$-$9000\,\AA. LAMOST spectra are released in consecutive data releases (DRs) and have been successfully employed in the study of CP stars in the past \citep{hou15,qin19,huemmerich20,paunzen21,shang22,huemmerich22,shi23,tian23}.

\begin{figure}
 \includegraphics[width=\columnwidth]{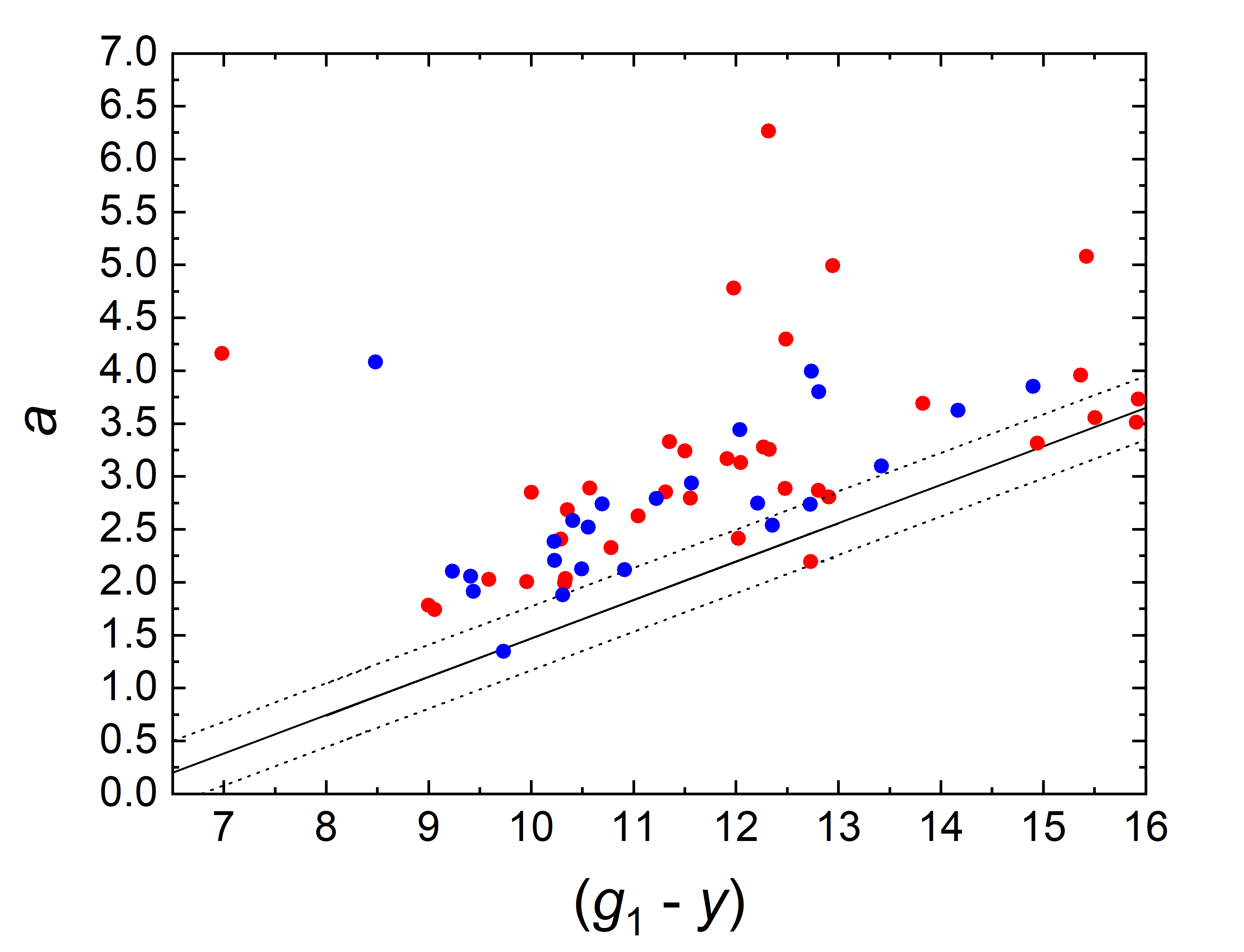}
    \caption{The $a$ versus ($g_1-y$) diagram for the 38 stars from our sample (red) and the 24 stars from the list of \citet[][blue]{mathys24} that have Gaia BP/RP spectra available. The normality line $a = -2.16(8) + 0.363(8) (g_1-y)$ is defined as in the classical $\Delta$a photometric system. The dotted lines are the 95\% prediction bands to select mCP stars.}
    \label{fig:Delta_a}
\end{figure}

\begin{figure*}
    \includegraphics[width=1.00\textwidth]{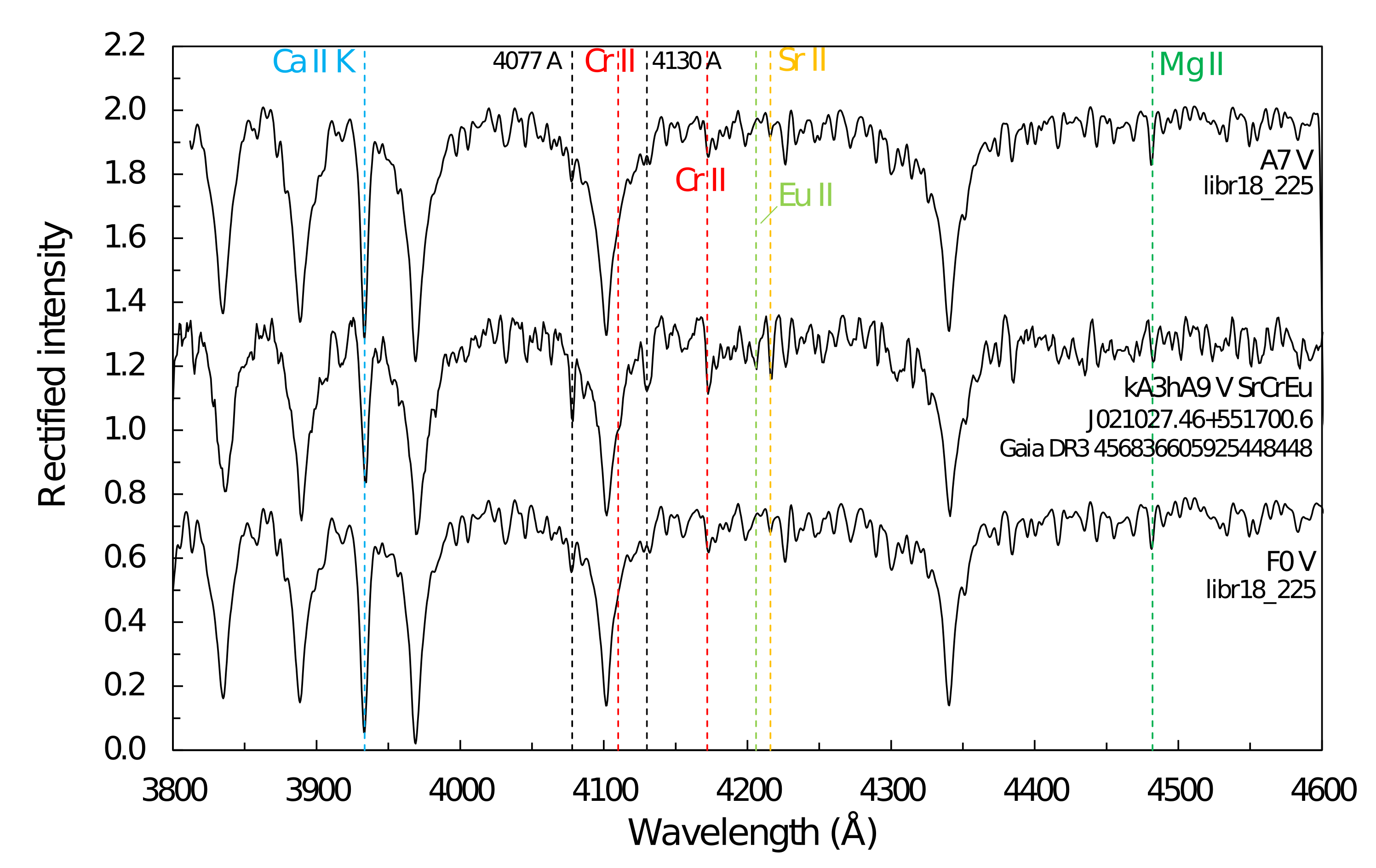}
    \caption{Blue-violet region of the LAMOST spectrum of the Ap star Gaia DR3 456836605925448448 = LAMOST J021027.46+551700.6 (middle spectrum), compared with two standard star spectra taken from the \textit{libr18\_225} collection. Some prominent lines and blends relevant to the classification of Ap stars are identified. We note the peculiarly strong \ion{Cr}{ii}, \ion{Sr}{ii}, and \ion{Eu}{ii} features and the weak \ion{Ca}{ii} K line in the Ap star.}
    \label{fig:showcase_spectra}
\end{figure*}

\section{Methodology}
\label{sect:methods}

This section details the methods used to identify, classify and investigate the new sample of ssrAp stars. 

\subsection{Candidate selection}
\label{sect:target_selection}

\citet{chen20} employed ZTF DR 2 to compile a catalogue of 781\,602 periodic variable stars, which is hereafter referred to as the ``ZTF catalogue of periodic variables''. They also compiled a catalogue of suspected variables stars (hereafter ``ZTF catalogue of suspected variables'') that contains more than 1\,300\,000 entries but only lists basic data and no classifications.

The ZTF catalogue of periodic variables contains classificatory information but lacks a specific category for ACV variables, which were consequently wrongly listed under other categories. This has been shown by \citet{faltova21} and \citet{bf24}, who identified samples of ACV variables that were incorrectly assigned to the class of RS Canum Venaticorum (RS CVn) stars in the ZTF catalogue of periodic variables. RS CVn stars are rotational variables whose light curves are superficially similar to the light curves of ACV variables. They are, however, a very different group of objects that consists of close binary stars with a giant component, active chromospheres, and enhanced spot activity \citep[e.g.][]{hall76}, and can be easily distinguished from ACV variables by for example the use of a colour-magnitude or Hertzsprung-Russell diagram (HRD) and an investigation of the light curve because ACVs show stable spot configurations (``chemical spots'') whereas the spots on RS CVn stars are prone to change (classical ``temperature spots''; \citealt{phillips24}).

To identify ACV candidates in the ZTF catalogue of periodic variables, \citet{faltova21} and \citet{bf24} used the following criteria, which are based on known characteristics of ACV variables \citep[e.g.][]{netopil17,jagelka19}: (a) variability period between one and ten days; (b) amplitude in the ZTF $r$ band of less than 0.3 mag; (c) the presence of a single independent variability frequency and corresponding harmonics; (d) stable or marginally changing light curve throughout the covered time span; and (e) an effective temperature between 6000 K and 25000 K \citep{andrae18}. As the above mentioned studies only targeted stars with rotation periods of $P$\,$\le$\,10\,d (item (a)), any ACV variables with periods in excess of that limit -- and, hence, any ssrAp stars ($P$\,$\ge$\,50\,d) -- will have been missed.

To specifically search for ssrAp stars, we therefore modified item (a) to ``variability period in excess of 50 days''. This, however, precluded a search for candidates among the RS CVn stars because \citet{chen20} applied a period cut-off of $P$\,$<$\,20\,d to this class of variable stars. For that reason, we had to search for an alternative starting point and chose to expand the search for new ACV variables in the ZTF catalogue of periodic variables to the class of semiregular (SR) variables. This class was chosen because the inclusion criteria of \citet[][variability period in excess of 20 days, amplitude of less than 2 mag]{chen20} do not include effective temperature or spectral type. We therefore expected that, if present, any ssrAp stars would have been assigned to this class. In addition, we also searched the ZTF catalogue of suspected variables for suitable candidates. For the effective temperature estimates, we opted to consult the more recent catalogue of \citet{anders22} instead of the compilation by \citet{andrae18} which was used by \citet{faltova21} and \citet{bf24}.

In summary, the candidate search for new ACV variables in the \citet{chen20} catalogues was conducted using the following criteria: (a) variability period > 50 days; (b) amplitude in the ZTF $r$ band of less than 0.3 mag; (c) presence of a single independent variability frequency and corresponding harmonics; (d) stable or marginally changing light curve throughout the covered time span; (e) effective temperature between 6000 K and 25 000 K, taken from the newer compilation of \citet{anders22}. Items (c), (d), and (e) are helpful to distinguish ACV variables from classical semiregular stars, which are not strictly periodic and generally cool red or yellow giants. 450 stars satisfied these criteria, which were then classified using photometric and spectroscopic properties, as detailed in the following sections.

\subsection{Photometric classification}
\label{sect:photometric_classification}

\subsubsection{ACV photometric properties}
\label{subsect:classification_photprop}
The light curves of all candidates were downloaded from the ZTF website\footnote{\url{https://www.ztf.caltech.edu}} and visually inspected to search for the typical light curve shape and photometric peculiarities of ACV variables and sort out contaminating objects. In particular, we searched for the following properties that have been shown to be characteristic of ACV variables \citep{groebel17,faltova21,bf24}: (i) ZTF $g$ and $r$ light curves are in anti-phase; (ii) the amplitude in $r$ is larger than the amplitude in $g$; and (iii) the amplitude in $r$ is about the same as the amplitude in $g$. Only one of these criteria needs to be fulfilled for a given object to be considered an ACV variable.

In combination with the pre-selection process (items (a)$-$(e); cf. Section \ref{sect:target_selection}), these items allow for the reliable identification of ACV variables in the absence of spectroscopic data.  Pulsating variables, for instance, are generally not strictly monoperiodic and will have been effectively excluded via candidate search item (c). Any remaining monoperiodic long-period pulsators, such as very long-period Cepheids, are sorted out via items (ii) and (iii) because pulsating variables are expected to show larger amplitudes at shorter wavelengths \citep[\textit{e.g.}][]{soszynski24}. The light curves of ellipsoidal variables may be particularly hard to tell apart from the light curves of ACV variables. However, only ellipsoidal variables with evolved components may exhibit periods in the targeted range. Any such objects -- as well as any evolved rotational variables such as RS CVn stars -- are easily identified by their outlying positions in the Hertzsprung-Russell diagram. In summary, no other known objects in the investigated temperature realm exhibit this particular combination of photometric properties and long-term stable monoperiodic variability \citep{groebel17,faltova21,bf24}. In this way, 392 candidates were discarded because the variability was spurious or none of the items (i)$-$(iii) was applicable. The remaining 58 stars were retained.

In the same manner, an additional search for ssrAp stars was carried out in ZTF raw data, which was downloaded for 10 percent (0-36°) of the sky area captured by ZTF and analysed for significant periods using the Lomb-Scargle algorithm as implemented in the program package \textsc{PERANSO} \citep{PERANSO}. Due to its extent, not all of the raw data could be processed. About 66\,100 objects were analysed in this way, which resulted in the discovery of 12 additional ssrAp stars. In summary, a final sample of 70 ssrAp stars was obtained.

\begin{figure}
    \includegraphics[width=\columnwidth]{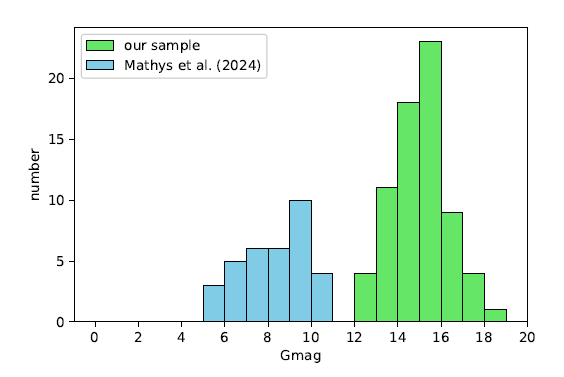}
    \caption{Distribution of $G$ magnitude brightness among our target star sample (green) and the sample of known ssrAp stars from \citet[][blue]{mathys24}.}
    \label{fig:histogramsGmag}
\end{figure}

\subsubsection{$\Delta$a photometry}
\label{subsect:classification_deltaa}

The $\Delta$a photometry tool is potent in investigating CP stars \citep{2005A&A...441..631P}. It investigates the flux depression at 5200\,\AA, a spectral feature
solely occurring in CP stars. The photometric system samples
the depth of this feature by comparing the flux at the centre with the adjacent regions using bandwidths of 
110 to 230\,\AA. 
The flux depression is caused by line blanketing of Cr, Fe, and Si enhanced by a magnetic
field \citep{2003MNRAS.341..849K,2007A&A...469.1083K}. It is most significantly visible in magnetic CP stars.
However, not all these objects show this flux depression, 
probably because of observational reasons (an unfavourable inclination) and the magnetic 
field characteristics \citep{2005A&A...441..631P}. In addition, some (non- or only weakly magnetic)
CP1 and CP3 stars also show detectable positive $\Delta$a values but with much less significance.

We used the approach presented by \citet{2022A&A...667L..10P} who used $Gaia$ BP/RP spectra to synthesize $\Delta$a photometry. They found a detection level of more
than 85\% for almost the entire investigated spectral range of the upper main sequence. 38 of our sample stars have a BP/RP spectrum available. According to \citet{2022A&A...667L..10P},
a $\Delta$a value larger than 0.3 is significant. 31 sample stars fulfil this criterion (Table \ref{table_parameters_Deltaa}), which is in line with the detection level of $\sim$85\% and additional independent proof that these objects are bona-fide mCP stars. The remaining 7 objects have no significant detectable flux depression at 5200\,\AA\ (which does not a priori exclude them from being Ap stars).

The $\Delta$a values of our sample stars are listed in Table \ref{table_parameters_Deltaa}, together with $\Delta$a values for the 24 known ssrAp stars from Table 1 of \citet{mathys24} that also have a BP/RP spectrum available. The $a$ versus ($g_1-y$) diagram for all these objects is shown in Figure \ref{fig:Delta_a}. The $\Delta$a values for the \citet{mathys24} stars were determined in the same way as described above.

\subsection{Spectral classification} \label{sect:spectral_classification}

All available low-resolution spectra for our sample stars were downloaded from the LAMOST spectral archive.\footnote{\url{http://www.lamost.org}} Only spectra of sufficient signal-to-noise ratio (SNR\,$\ge$\,20) were considered. In this way, spectra were downloaded for five sample stars.

Spectral classification was performed in the framework of the refined MK classification system following \citet{gray87,gray89a,gray89b}, \citet{gray94} and \citet{gray09}. To derive a precise classification and identify peculiarities, the blue-violet (3800$-$4600\,\AA) spectral region was compared visually to, and overlaid with, MK standard star spectra, which were taken from the \textit{libr18} collections available from R. O. Gray's MKCLASS website.\footnote{\url{http://www.appstate.edu/~grayro/mkclass/}} Where appropriate, spectral types based on the \ion{Ca}{ii} K line strength (the k-line type) and the hydrogen-line profile (the h-line type) were determined \citep{osawa65,gray09}. As the metallic lines of most Ap stars are so peculiar that they cannot be used for luminosity classification, luminosity types were based on the wings of the hydrogen lines \citep{gray09}.

An example of this process is illustrated in Figure \ref{fig:showcase_spectra}. All stars classified in this way turned out to be bona-fide Ap stars, which is again independent proof of the efficiency of the sample selection process. In addition, all spectra showed the characteristic 5200\,\AA\ depression.

\begin{figure*}[h!]
    \includegraphics[width=\textwidth]{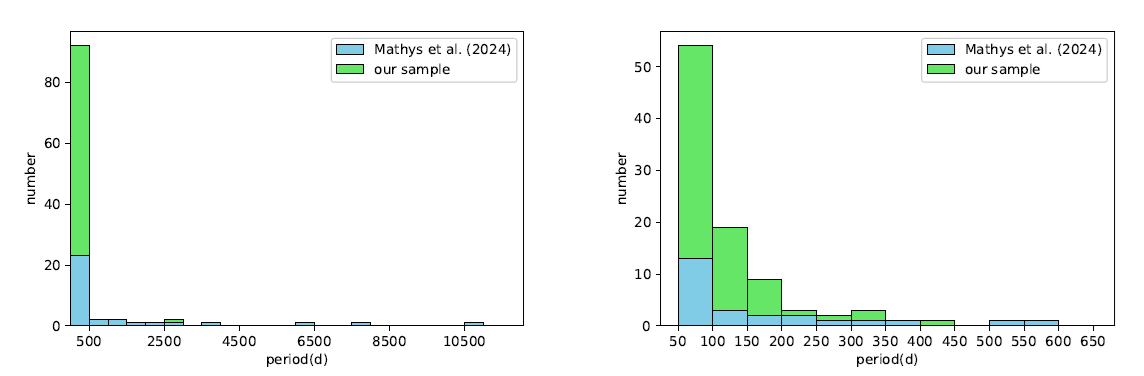}
    \caption{Distribution of rotational periods among our target star sample (green) and the sample of known ssrAp stars from \citet{mathys24} (blue). The right panel provides a zoom-in on the period range from 50 to 650 days.}
    \label{fig:histogramsProt}
\end{figure*}

\begin{figure}
    \includegraphics[width=\columnwidth]{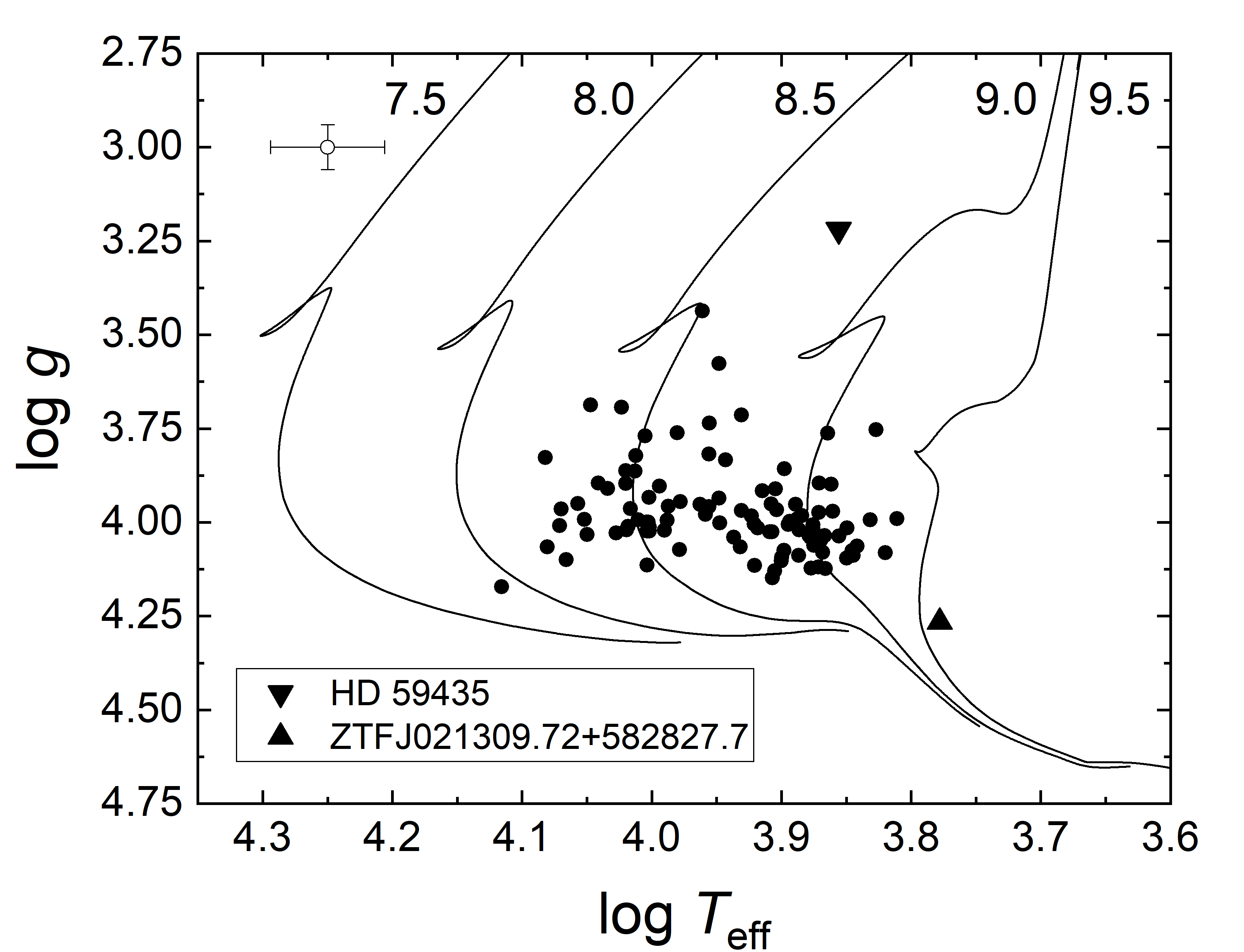}
    \caption{The HRD for our targets stars together with the PARSEC isochrones \citep{2012MNRAS.427..127B} for the listed logarithmic ages and solar metallicity [Z]\,=\,0.02\,dex. In the left corner are the mean errors for the individual data shown.
    The two outliers (HD 59435 and ZTFJ021309.72+582827.7) are discussed in the text.}
    \label{fig:HRD}
\end{figure}

\section{Results}
\label{sect:results}

\subsection{Basic parameters}

Basic parameters of the final sample of 70 ssrAp stars identified in this study are listed in the appendix in Table \ref{table_results1}. Positional information and magnitude in the $G$\,band were taken from Gaia DR3 \citep{GAIA1,GAIA2,GAIA3}. All periods were derived using the Lomb-Scargle algorithm from PERANSO (\citealt{PERANSO}; cf. also Section \ref{sect:target_selection}). The abbreviations in column four indicate the following sources: ``ZTF\_per'':  ZTF catalogue of periodic variables \citep{chen20}; ``ZTF\_sus'': ZTF catalogue of suspected variables \citep{chen20}; ``raw data'': ZTF raw data. Column eight lists the main photometric property the star was selected for: ``antiphase'': the $g$ and $r$ light curves are in anti-phase; ``amp $r>g$'': the amplitude of the variations in $r$ is larger than the amplitude in $g$; ``amp $r\sim g$'': the amplitude of the variations in $r$ is about the same as the amplitude in $g$; ``LC shape'':  characteristic light curve shape.

Figure \ref{fig:histogramsGmag} investigates the distribution of $G$ magnitudes for our sample stars and the sample of ssrAp stars from \citet{mathys24}, who presented a census of the presently known ssrAp stars based on accurately determined period values or established period lower limits. The magnitude distribution of our sample stars has a broad maximum at $G$ $\sim$ 15 mag (mean magnitude $G$ = 15.05 mag). Our work, therefore, extends the sample of known ssrAp stars to significantly lower magnitudes.

Figure \ref{fig:histogramsProt} investigates the distribution of rotational periods for the ssrAp stars from our sample and the sample of \citet{mathys24}. For clarity, objects with period lower limits from the latter source are not included in the plot. It is obvious that our stars are situated at the lower period end of the period distribution of ssrAp stars that shows a long tail petering out towards rotational periods of thousands of days and more. 42 of our sample stars have periods of 50\,$\le P\textsubscript{rot} \le$\,100 days, another 22 stars show periods of 100\,$< P\textsubscript{rot} \le$\,200 days. Only six stars show longer periods, with the longest-period object (Gaia DR3 456836605925448448) exhibiting a rotational period of P\textsubscript{rot} = 2551.7 days). This is to be expected: due to the limited time coverage, ZTF data are not well suited to the discovery of ssrAp stars with very long periods.

Phase plots for all our sample stars are provided in the appendix in Figure \ref{fig:phaseplots1}.

\subsection{Astrophysical parameters and Hertzsprung-Russell Diagram}

\label{sect:astropyhsical_parameters}

To locate the target stars in the HRD, we calibrated effective temperature ($T_\mathrm{eff}$) and surface gravity ($\log g$) using
the calibration published by \citet{2024A&A...683L...7P}. The latter is based
on astrophysical parameters automatically determined by four independent methods using
photometric and spectroscopic data. The paper presents
a statistical analysis comparing the $T_\mathrm{eff}$ and $\log g$ from high-resolution spectroscopy and
the sources correcting for offsets. 

Table \ref{table_parameters_Deltaa} presents
the mean values and their errors. The mean standard deviation of all $T_\mathrm{eff}$
values is 10.4\% whereas it is 2.2\% for $\log g$, respectively.
Figure \ref{fig:HRD} shows the HRD together 
with PARSEC isochrones \citep{2012MNRAS.427..127B} for solar metallicity [Z]\,=\,0.02\,dex.
As reported before \citep{huemmerich20,paunzen21}, the target stars are, with very few
exceptions, grouped
in the middle of the main sequence with ages older than 150\,Myr. 

\begin{figure}
\includegraphics[width=\columnwidth]{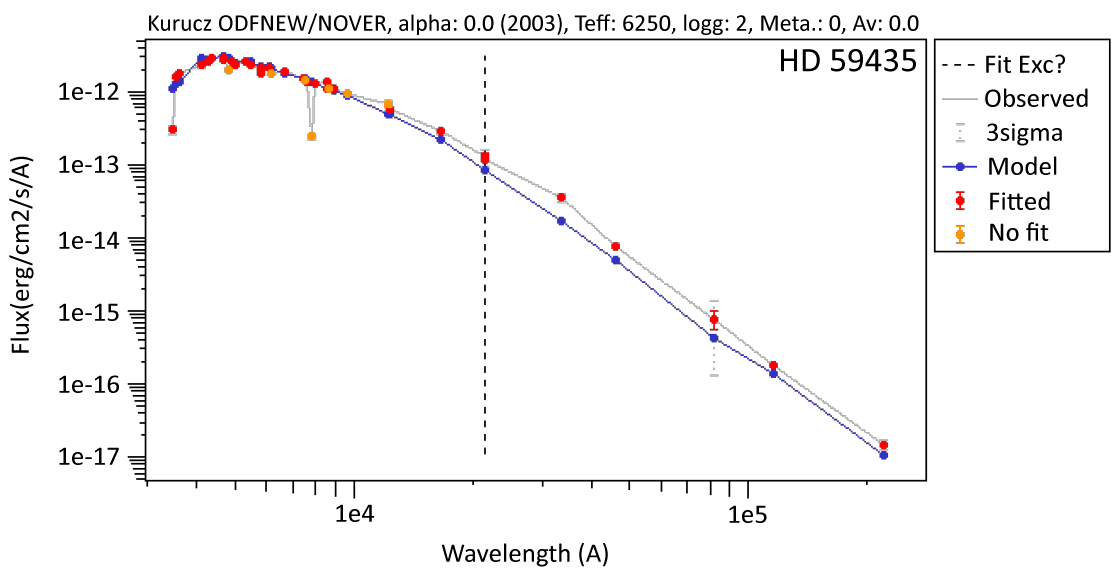} \includegraphics[width=\columnwidth]{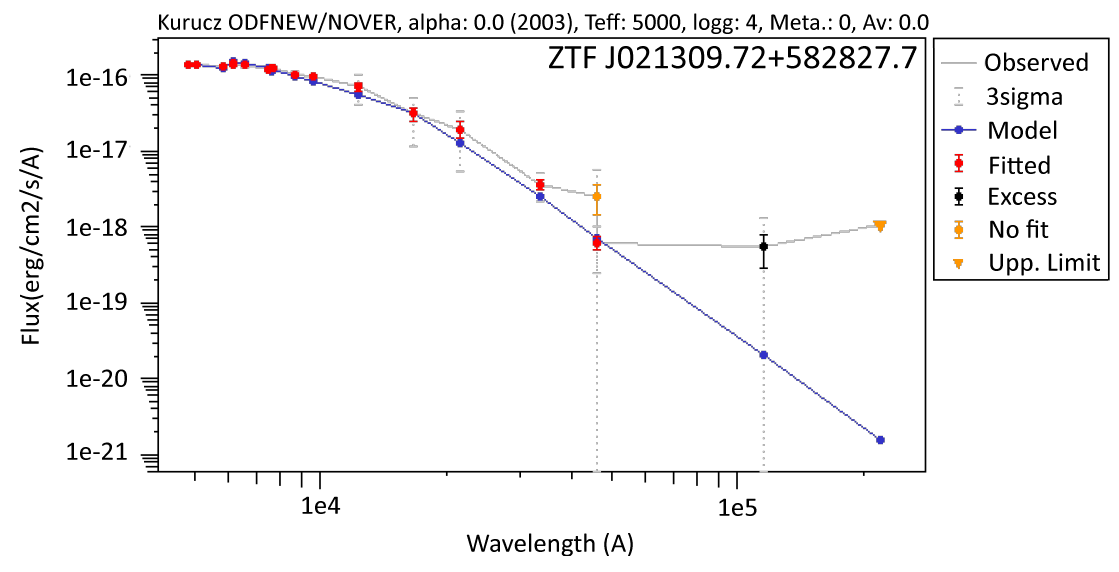}
\caption{SED plots for HD 59435 (upper panel) and ZTF J021309.72+582827.7 (lower panel), which were created with the VOSA tool.}
\label{fig:SED}
\end{figure}

Two stars are conspicuous by their outlying positions in the HRD and deserve special mention. HD 59435 is a well-established ssrAp star from the list of \citet{mathys24}. It is a non-eclipsing double-lined spectroscopic binary (SB2) system consisting of a G8/K0 primary star and a cool SrCrEu (ssr)Ap secondary star \citep{wade96}. The presence of the giant star explains its outlying position in the HRD and is also clearly visible in the SED plot (Fig. \ref{fig:SED}, upper panel), which was created with the VO SED Analyzer (VOSA; \citealt{bayo08}), which is a browser-based online tool.

ZTF J021309.72+582827.7, on the other hand, is a ssrAp star from the new sample. Assuming a main-sequence object, the derived temperature of $\sim$6000 K would put this star at around a spectral type of F9, which is much too late for a classical Ap star. At this temperature, convection is expected to mix the stellar atmosphere, which effectively prevents the establishment of surface chemical peculiarities. Therefore, for main-sequence stars later than about F5, the typical Ap star peculiarities are not observed \citep{gray09}. There are, however, clear indications that ZTF J021309.72+582827.7 is indeed a chemically peculiar object. First, the $\Delta a$ value of 1.923 mag is a clear detection. Second, the amplitude of the photometric variations in $r$ is much larger than the amplitude in $g$: while the star shows a ``healthy'' amplitude of about 0.2\,mag in the $r$ band, the variations in the $g$ band are hardly noticeable with an amplitude of only about 0.05\,mag. There are also slight indications of an antiphase pattern, that is, when the star is brighter in $r$, it is fainter in $g$ and vice versa (cf. Fig. \ref{fig:phaseplots1}). The SED plot (Fig. \ref{fig:SED}, lower panel) indicates a strong infrared (IR) excess. Perhaps, ZTF J021309.72+582827.7 is a binary star harbouring an Ap star and a cool component, possibly shrouded in dust. The strong IR excess could also point to a young stellar object; there is, however, no indication that the star is situated in a region of the sky with ongoing star formation. Only further spectroscopic investigations will be able to shed more light on this interesting object's true nature.

\section{Conclusion}
\label{sect:conclusion}

We have carried out a search for new ssrAp stars using data from the ZTF survey. Using suitable selection criteria based on known astrophysical and light curve properties of this group of stars and, where available, $\Delta$a photometry and LAMOST spectra, we identified 70 new ssrAp stars. With a mean magnitude of $G$\,$\sim$\,15\,mag, we extend the sample of known ssrAp stars to significantly lower magnitudes. Our sample stars mostly exhibit rotational periods between 50 and 200 days, with the longest period object having a period of P\textsubscript{rot} = 2551.7 days (Gaia DR3 456836605925448448). We stress that because of the limited time span of the ZTF data, only Ap stars with rotational periods of up to about 3000 d could be unambiguously identified in our study which, therefore, is not suited to detect Ap stars with very long rotation periods.

We have presented astrophysical parameters and a HRD for the sample of new ssrAp stars and the sample of known ssrAp stars with period determinations from \citet{mathys24}. With only a few exceptions, all ssrAp stars are grouped in the middle of the main sequence with ages older than 150 Myr. Furthermore, ZTF J021309.72+582827.7 was identified as a possible binary star harbouring an Ap star and a cool component, possibly shrouded in dust, thus being a prime candidate for further detailed investigations.

With our work, we enlarge the sample of known ssrAp stars by about 150\%, thereby paving the way for more in-depth statistical studies. Further investigations of our sample stars should include spectroscopic observations and magnetic field measurements.

\begin{acknowledgements}
We thank the referee, Dr. Gautier Mathys, for his valuable comments that improved the paper. We also thank Prof. Dr. Nikolay N. Samus for his help to identify the references pertaining to the initial denomination of the class of ACV variables. This study is based on observations obtained with the Samuel Oschin 48-inch Telescope at the Palomar Observatory as part of the Zwicky Transient Facility project, which is supported by the National Science Foundation under Grant No. AST-1440341 and a collaboration including Caltech, IPAC, the Weizmann Institute for Science, the Oskar Klein Center at Stockholm University, the University of Maryland, the University of Washington, Deutsches Elektronen-Synchrotron and Humboldt University, Los Alamos National Laboratories, the TANGO Consortium of Taiwan, the University of Wisconsin at Milwaukee, and Lawrence Berkeley National Laboratories. Operations are conducted by COO, IPAC, and UW. This work also uses data from the Guoshoujing Telescope (the Large Sky Area Multi-Object Fiber Spectroscopic Telescope LAMOST), which is a National Major Scientific Project built by the Chinese Academy of Sciences. Funding for the project has been provided by the National Development and Reform Commission. LAMOST is operated and managed by the National Astronomical Observatories, Chinese Academy of Sciences. In addition, this work has made use of data from the European Space Agency (ESA) mission {\it Gaia} (\url{https://www.cosmos.esa.int/gaia}), processed by the {\it Gaia} Data Processing and Analysis Consortium (DPAC, \url{https://www.cosmos.esa.int/web/gaia/dpac/consortium}). Funding for the DPAC has been provided by national institutions, in particular the institutions participating in the {\it Gaia} Multilateral Agreement.  
\end{acknowledgements}

%
%
\bibliographystyle{aa}
\bibliography{ssrAp.bib}

\begin{appendix} 

\section{Tables}

\begin{table*}
\caption{Astrophysical parameters and synthetic $\Delta a$ photometry for the merged samples of ssrAp stars and candidates.}  
\tiny
\label{table_parameters_Deltaa}
\begin{center}
\begin{adjustbox}{max width=1.4\textwidth,angle=90}
\begin{tabular}{lllccccc|lllccccc}
\hline
\hline
(1) & (2) & (3) & (4) & (5) & (6) & (7) & (8) & (1) & (2) & (3) & (4) & (5) & (6) & (7) & (8) \\
\hline
ZTFJ000301.53+594058.8	&	7848	&	469	&	4.005	&	0.040	&		&		&		&	HD 93507	&	10247	&	648	&	3.993	&	0.095	&	9.411	&	2.057	&	+0.800	\\
ZTFJ000551.46+623437.4	&	9034	&	1183	&	3.958	&	0.065	&		&		&		&	HD 94660	&	10816	&	468	&	3.910	&	0.054	&	11.565	&	2.938	&	+0.900	\\
Gaia DR3 431305499257132032	&	9519	&	1422	&	4.073	&	0.122	&		&		&		&	HD 102797	&	10470	&	1343	&	3.863	&	0.117	&	10.558	&	2.519	&	+0.846	\\
HD 965	&	7435	&	276	&	3.974	&	0.022	&	12.737	&	3.994	&	+1.531	&	HD 103844	&	10043	&	1898	&	4.023	&	0.127	&	10.493	&	2.126	&	+0.477	\\
ZTFJ001856.17+645001.5	&	10084	&	1137	&	4.024	&	0.091	&		&		&		&	TYC 8979-339-1	&	7511	&	478	&	4.018	&	0.053	&	10.407	&	2.582	&	+0.965	\\
Gaia DR3 430596245536174080	&	8872	&	123	&	3.577	&	0.125	&		&		&		&	HD 110274	&	8014	&	323	&	3.967	&	0.036	&	14.169	&	3.623	&	+0.640	\\
HD 2453	&	9175	&	217	&	3.953	&	0.020	&	12.039	&	3.441	&	+1.231	&	HD 116458	&	10048	&	261	&	3.933	&	0.020	&	13.419	&	3.098	&	+0.386	\\
ZTFJ003714.92+611125.0	&	8379	&	891	&	3.983	&	0.033	&	12.480	&	2.886	&	+0.515	&	HD 123335	&	11147	&	1566	&	3.687	&	0.229	&	10.227	&	2.383	&	+0.830	\\
Gaia DR3 417652416694261120	&	7499	&	217	&	4.061	&	0.072	&	12.946	&	4.993	&	+2.454	&	HD 126515	&	10087	&	173	&	4.114	&	0.151	&	9.732	&	1.345	&	$-$0.027	\\
ZTFJ005521.62+710055.4	&	7076	&	1054	&	4.096	&	0.092	&	15.928	&	3.730	&	+0.108	&	HD 149766	&	10121	&	2259	&	3.770	&	0.058	&		&		&		\\
ZTFJ005643.06+641256.7	&	7544	&	774	&	4.040	&	0.088	&		&		&		&	HD 158919	&	10432	&	1033	&	4.011	&	0.102	&	10.228	&	2.205	&	+0.652	\\
HD 5797	&	9031	&	660	&	3.735	&	0.169	&		&		&		&	HD 166473	&	8041	&	315	&	4.129	&	0.121	&	10.693	&	2.741	&	+1.019	\\
Gaia DR3 426587872814713216 	&	11402	&	1185	&	3.950	&	0.022	&	11.913	&	3.167	&	+1.003	&	ZTFJ190052.47+025113.7	&	7429	&	543	&	3.896	&	0.049	&	14.942	&	3.314	&	+0.050	\\
Gaia DR3 522541599739803008	&	9862	&	2973	&	3.904	&	0.089	&	10.355	&	2.684	&	+1.085	&	ZTFJ191752.71+201804.0	&	8072	&	328	&	4.148	&	0.171	&		&		&		\\
HD 8441	&	9143	&	374	&	3.437	&	0.000	&	9.436	&	1.913	&	+0.648	&	ZTFJ192438.19+755932.0	&	7472	&	607	&	4.035	&	0.097	&		&		&		\\
ZTFJ012628.29+615404.2	&	11272	&	2242	&	3.993	&	0.057	&	10.334	&	2.034	&	+0.443	&	ZTFJ193202.56-001234.9	&	7252	&	51	&	3.971	&	0.021	&	15.504	&	3.554	&	+0.086	\\
HD 9996	&	10063	&	362	&	3.999	&	0.040	&	8.482	&	4.081	&	+3.162	&	HD 184471	&	7900	&	401	&	3.857	&	0.133	&		&		&		\\
Gaia DR3 509947140719557120	&	10040	&	1956	&	4.013	&	0.139	&	10.782	&	2.327	&	+0.573	&	ZTFJ193723.37+225923.5	&	6717	&	1300	&	3.753	&	0.006	&	12.728	&	2.193	&	$-$0.267	\\
ZTFJ014032.09+624144.2	&	9032	&	111	&	3.818	&	0.124	&	11.978	&	4.779	&	+2.591	&	ZTFJ194031.94+155946.6	&	11774	&	2110	&	4.009	&	0.068	&		&		&		\\
Gaia DR3 409362030294843648	&	7410	&	693	&	4.048	&	0.095	&	10.572	&	2.889	&	+1.211	&	ZTFJ194349.09+381543.7	&	7828	&	330	&	3.998	&	0.039	&	9.060	&	1.741	&	+0.612	\\
Gaia DR3 511800569419741056	&	11641	&	2443	&	4.100	&	0.180	&		&		&		&	HD 187474	&	10384	&	104	&	3.964	&	0.006	&	12.724	&	2.735	&	+0.276	\\
Gaia DR3 358029955561056896	&	7700	&	634	&	4.020	&	0.080	&	11.502	&	3.242	&	+1.227	&	HD 188041	&	8529	&	226	&	3.969	&	0.016	&		&		&		\\
ZTFJ015319.49+635118.4	&	11748	&	2900	&	3.965	&	0.124	&	15.420	&	5.079	&	+1.642	&	ZTFJ195518.79+295545.9	&	8338	&	2069	&	4.005	&	0.093	&		&		&		\\
ZTFJ015931.47+632340.7	&	12084	&	4061	&	3.827	&	0.131	&	9.958	&	2.006	&	+0.552	&	ZTFJ201036.73+365925.3	&	9090	&	1435	&	3.979	&	0.012	&		&		&		\\
Gaia DR3 456836605925448448	&	7562	&	387	&	4.036	&	0.067	&	9.000	&	1.781	&	+0.674	&	HD 340577	&	10301	&	2155	&	3.864	&	0.051	&		&		&		\\
Gaia DR3 506670183748373760	&	8113	&	356	&	4.026	&	0.083	&	12.324	&	3.255	&	+0.942	&	HD 200311	&	12038	&	172	&	4.066	&	0.005	&	12.354	&	2.538	&	+0.213	\\
ZTFJ021309.72+582827.7	&	5997	&	962	&	4.266	&	0.129	&	12.487	&	4.296	&	+1.923	&	ZTFJ210819.95+482948.2	&	13054	&	5098	&	4.171	&	0.075	&	9.589	&	2.025	&	+0.705	\\
Gaia DR3 456715453486872192	&	7909	&	154	&	4.075	&	0.091	&		&		&		&	ZTFJ211053.69+432024.2	&	7275	&	847	&	3.899	&	0.035	&		&		&		\\
HD 18078	&	10475	&	2353	&	3.897	&	0.120	&	12.213	&	2.748	&	+0.475	&	ZTFJ212540.46+594620.2	&	7071	&	989	&	4.015	&	0.077	&		&		&		\\
ZTFJ034324.66+521241.2	&	9509	&	955	&	3.945	&	0.033	&	11.315	&	2.852	&	+0.905	&	ZTFJ212623.27+551353.6	&	8530	&	1698	&	3.714	&	0.073	&	12.316	&	6.265	&	+3.954	\\
ZTFJ034958.11+563711.7	&	7940	&	413	&	4.093	&	0.104	&	6.981	&	4.160	&	+3.786	&	ZTFJ213025.59+475535.3	&	8876	&	418	&	3.935	&	0.039	&	8.956	&	9.774	&	+8.683	\\
ZTFJ044805.24+481447.1	&	7947	&	218	&	4.102	&	0.142	&		&		&		&	ZTFJ213425.23+504604.4	&	7009	&	764	&	4.076	&	0.099	&	11.351	&	3.327	&	+1.367	\\
ZTFJ050617.31+490300.5	&	8220	&	318	&	3.916	&	0.112	&		&		&		&	ZTFJ213435.94+595813.4	&	7512	&	476	&	4.007	&	0.048	&		&		&		\\
ZTFJ051811.15+401738.0	&	7172	&	695	&	4.037	&	0.081	&	12.267	&	3.276	&	+0.983	&	ZTFJ214050.19+521625.7	&	7747	&	653	&	3.952	&	0.025	&	12.047	&	3.132	&	+0.919	\\
ZTFJ053102.53+301449.3	&	8333	&	961	&	4.115	&	0.125	&		&		&		&	ZTFJ214501.59+643441.9	&	6783	&	1102	&	3.993	&	0.059	&		&		&		\\
ZTFJ054532.92+344429.5	&	8088	&	406	&	3.951	&	0.037	&		&		&		&	ZTFJ215742.24+492955.8	&	7705	&	81	&	4.088	&	0.096	&		&		&		\\
ZTFJ060010.70+151059.5	&	7537	&	787	&	4.122	&	0.113	&		&		&		&	ZTFJ215751.18+562522.6	&	8771	&	478	&	3.834	&	0.086	&	15.365	&	3.956	&	+0.539	\\
ZTFJ061054.19+080816.5	&	6991	&	914	&	4.088	&	0.081	&		&		&		&	ZTFJ220344.78+460557.5	&	7439	&	767	&	4.119	&	0.171	&		&		&		\\
HD 256476	&	11215	&	1844	&	4.033	&	0.134	&		&		&		&	ZTFJ221316.28+471544.8	&	7384	&	391	&	4.080	&	0.098	&		&		&		\\
ZTFJ062945.32+235829.3	&	7316	&	590	&	3.762	&	0.159	&	10.291	&	2.407	&	+0.831	&	ZTFJ221537.26+513506.5	&	9732	&	866	&	3.994	&	0.054	&		&		&		\\
HD 263361	&	10548	&	210	&	3.694	&	0.000	&	14.900	&	3.851	&	+0.603	&	ZTFJ223955.81+582547.7	&	8026	&	468	&	3.910	&	0.021	&	11.555	&	2.793	&	+0.759	\\
ZTFJ064705.05-132227.3	&	6607	&	797	&	4.081	&	0.076	&	10.327	&	1.994	&	+0.405	&	ZTFJ224236.86+521540.8	&	8546	&	315	&	4.065	&	0.072	&	10.003	&	2.849	&	+1.378	\\
HD 50169	&	10096	&	923	&	3.999	&	0.053	&		&		&		&	ZTFJ224614.86+540241.8	&	7723	&	402	&	3.991	&	0.025	&	12.022	&	2.415	&	+0.211	\\
HD 51684	&	8284	&	533	&	4.015	&	0.114	&	11.222	&	2.792	&	+0.878	&	ZTFJ225959.83+610427.5	&	7356	&	323	&	4.036	&	0.065	&	13.826	&	3.691	&	+0.832	\\
ZTFJ070734.93-063518.4	&	6469	&	738	&	3.991	&	0.034	&	12.906	&	2.804	&	+0.279	&	ZTFJ230106.58+591736.7	&	7662	&	613	&	3.982	&	0.031	&		&		&		\\
TYC 5395-1139-1	&	9555	&	559	&	3.761	&	0.157	&		&		&		&	ZTFJ230345.17+652605.3	&	8863	&	1265	&	4.002	&	0.075	&		&		&		\\
ZTFJ072235.79-062059.5	&	6947	&	542	&	4.064	&	0.093	&	11.044	&	2.625	&	+0.776	&	ZTFJ230923.96+645158.7	&	9711	&	1570	&	3.957	&	0.017	&		&		&		\\
HD 59435	&	7177	&	1033	&	3.220	&	0.616	&		&		&		&	ZTFJ232607.29+611952.2	&	7534	&	596	&	4.038	&	0.090	&		&		&		\\
HD 61468	&	9774	&	1056	&	4.021	&	0.047	&	12.808	&	3.798	&	+1.308	&	ZTFJ232932.44+612137.9	&	8078	&	235	&	4.025	&	0.064	&		&		&		\\
HD 66821	&	10656	&	1403	&	4.029	&	0.068	&	10.912	&	2.119	&	+0.318	&	HD 221568	&	10456	&	1751	&	4.020	&	0.117	&		&		&		\\
HD 69544	&	10286	&	1642	&	3.822	&	0.063	&	10.310	&	1.879	&	+0.297	&	ZTFJ235136.41+490754.2	&	7346	&	677	&	4.123	&	0.152	&	12.803	&	2.868	&	+0.381	\\
HD 87528	&	10995	&	1093	&	3.896	&	0.035	&	9.232	&	2.103	&	+0.912	&	ZTFJ235247.14+593239.3	&	8647	&	2564	&	4.040	&	0.078	&	15.909	&	3.510	&	$-$0.105	\\
\hline
\end{tabular} 
\end{adjustbox}
\tablefoot{The columns denote: (1) Identifier; (2) $T_\mathrm{eff}$; (3) $\sigma T_\mathrm{eff}$; (4) $\log g$; (5) $\sigma \log g$; (6) $(g1-y)$ colour; (7) $a$ index; (8) $\Delta a$ value.}
\end{center}   
\end{table*}

\begin{table*}
\caption{Basic parameters for the newly identified sample of ssrAp stars and candidates.}
\label{table_results1}
\begin{adjustbox}{max width=0.99\textwidth}
\begin{tabular}{lllllcll}
\hline
\hline
ID & RA (J2000) & Dec (J2000) & Source & $G$ mag & $P$ & Spectral type & Remark \\
\hline
ZTF J000301.53+594058.8	&	00 03 01.55	&	+59 40 58.79	&	ZTF\_per	&	15.901	&	74.7	&		&	antiphase	\\
ZTF J000551.46+623437.4	&	00 05 51.47	&	+62 34 37.40	&	ZTF\_per	&	15.841	&	70.9	&		&	amp $r>g$	\\
Gaia DR3 431305499257132032	&	00 13 06.77	&	+62 43 14.37	&	raw data	&	14.059	&	79.4	&		&	amp $r\sim g$	\\
ZTF J001856.17+645001.5	&	00 18 56.18	&	+64 50 01.53	&	ZTF\_per	&	15.259	&	61.2	&		&	amp $r>g$	\\
Gaia DR3 430596245536174080	&	00 23 32.57	&	+62 34 06.23	&	raw data	&	13.815	&	51.4	&		&	amp $r>g$	\\
ZTF J003714.92+611125.0	&	00 37 14.93	&	+61 11 24.94	&	ZTF\_per	&	14.356	&	140.2	&		&	antiphase	\\
Gaia DR3 417652416694261120	&	00 48 45.07	&	+55 06 58.41	&	raw data	&	13.968	&	144.3	&	kA3hA7 V SrCrEu	&	amp $r>g$	\\
ZTF J005521.62+710055.4	&	00 55 21.63	&	+71 00 55.30	&	ZTF\_per	&	17.053	&	51.1	&		&	antiphase	\\
ZTF J005643.06+641256.7	&	00 56 43.07	&	+64 12 56.67	&	ZTF\_per	&	16.025	&	77.7	&		&	antiphase	\\
Gaia DR3 426587872814713216 	&	01 02 53.09	&	+60 50 35.19	&	raw data	&	13.512	&	408.9	&		&	amp $r>g$	\\
Gaia DR3 522541599739803008	&	01 07 00.22	&	+61 28 26.32	&	raw data	&	13.809	&	66.6	&		&	LC shape	\\
ZTF J012628.29+615404.2	&	01 26 28.30	&	+61 54 04.10	&	ZTF\_per	&	14.432	&	177.5	&		&	amp $r\sim g$	\\
Gaia DR3 509947140719557120	&	01 38 53.80	&	+61 08 49.89	&	raw data	&	14.891	&	86.1	&		&	amp $r>g$	\\
ZTF J014032.09+624144.2	&	01 40 32.10	&	+62 41 44.18	&	ZTF\_per	&	13.928	&	54.7	&		&	amp $r\sim g$	\\
Gaia DR3 409362030294843648	&	01 41 34.97	&	+55 11 40.38	&	raw data	&	15.130	&	196.8	&		&	antiphase	\\
Gaia DR3 511800569419741056	&	01 49 34.73	&	+62 40 32.78	&	raw data	&	14.312	&	142.4	&		&	amp $r\sim g$	\\
Gaia DR3 358029955561056896	&	01 51 29.84	&	+51 01 35.64	&	raw data	&	14.821	&	322.3	&		&	antiphase	\\
ZTF J015319.49+635118.4	&	01 53 19.50	&	+63 51 18.36	&	ZTF\_per	&	13.926	&	113.3	&		&	amp $r>g$	\\
ZTF J015931.47+632340.7	&	01 59 31.48	&	+63 23 40.65	&	ZTF\_per	&	14.367	&	76.6	&		&	amp $r\sim g$	\\
Gaia DR3 456836605925448448	&	02 10 27.46	&	+55 17 00.59	&	raw data	&	13.767	&	2551.7	&	kA3hA9 V SrCrEu	&	antiphase	\\
Gaia DR3 506670183748373760	&	02 10 41.85	&	+57 50 17.75	&	raw data	&	14.422	&	139.9	&		&	antiphase	\\
ZTF J021309.72+582827.7	&	02 13 09.73	&	+58 28 27.70	&	ZTF\_per	&	18.211	&	60.8	&		&	amp $r>g$ \\
Gaia DR3 456715453486872192	&	02 15 54.98	&	+55 26 30.95	&	raw data	&	12.942	&	55.8	&		&	amp $r>g$	\\
ZTF J034324.66+521241.2	&	03 43 24.66	&	+52 12 41.24	&	ZTF\_per	&	14.656	&	59.2	&		&	amp $r\sim g$	\\
ZTF J034958.11+563711.7	&	03 49 58.12	&	+56 37 11.59	&	ZTF\_per	&	15.717	&	59.1	&		&	LC shape	\\
ZTF J044805.24+481447.1	&	04 48 05.25	&	+48 14 47.07	&	ZTF\_per	&	15.817	&	57.2	&		&	amp $r>g$	\\
ZTF J050617.31+490300.5	&	05 06 17.32	&	+49 03 00.48	&	ZTF\_per	&	14.519	&	64.2	&		&	amp $r>g$	\\
ZTF J051811.15+401738.0	&	05 18 11.15	&	+40 17 37.97	&	ZTF\_per	&	15.624	&	100.0	&		&	antiphase	\\
ZTF J053102.53+301449.3	&	05 31 02.54	&	+30 14 49.31	&	ZTF\_per	&	16.581	&	52.2	&		&	amp $r\sim g$	\\
ZTF J054532.92+344429.5	&	05 45 32.93	&	+34 44 29.42	&	ZTF\_per	&	15.997	&	64.8	&	A5: V: SrCrEu	&	antiphase	\\
ZTF J060010.70+151059.5	&	06 00 10.70	&	+15 10 59.54	&	ZTF\_per	&	16.406	&	272.5	&	B9 V SiSrCrEu	&	antiphase	\\
ZTF J061054.19+080816.5	&	06 10 54.20	&	+08 08 16.46	&	ZTF\_sus	&	16.293	&	300.6	&		&	antiphase	\\
ZTF J062945.32+235829.3	&	06 29 45.33	&	+23 58 29.12	&	ZTF\_per	&	14.873	&	67.2	&		&	antiphase	\\
ZTF J064705.05-132227.3	&	06 47 05.05	&	$-$13 22 27.40	&	ZTF\_sus	&	17.367	&	127.5	&		&	antiphase	\\
ZTF J070734.93-063518.4	&	07 07 34.94	&	$-$06 35 18.53	&	ZTF\_per	&	15.747	&	190.6	&		&	antiphase	\\
ZTF J072235.79-062059.5	&	07 22 35.79	&	$-$06 20 59.57	&	ZTF\_per	&	15.770	&	75.5	&		&	antiphase	\\
ZTF J190052.47+025113.7	&	19 00 52.48	&	+02 51 13.71	&	ZTF\_per	&	17.441	&	71.8	&		&	amp $r\sim g$	\\
ZTF J191752.71+201804.0	&	19 17 52.72	&	+20 18 04.07	&	ZTF\_per	&	15.819	&	75.3	&		&	antiphase	\\
ZTF J192438.19+755932.0	&	19 24 38.19	&	+75 59 32.11	&	ZTF\_per	&	14.269	&	50.3	&		&	antiphase	\\
ZTF J193202.56-001234.9	&	19 32 02.56	&	$-$00 12 35.05	&	ZTF\_per	&	14.004	&	79.7	&		&	antiphase	\\
ZTF J193723.37+225923.5	&	19 37 23.38	&	+22 59 23.66	&	ZTF\_per	&	16.229	&	111.2	&		&	amp $r\sim g$	\\
ZTF J194031.94+155946.6	&	19 40 31.95	&	+15 59 46.59	&	ZTF\_per	&	12.900	&	61.7	&		&	amp $r\sim g$	\\
ZTF J194349.09+381543.7	&	19 43 49.10	&	+38 15 43.69	&	ZTF\_per	&	13.854	&	93.9	&		&	antiphase	\\
ZTF J195518.79+295545.9	&	19 55 18.80	&	+29 55 45.94	&	ZTF\_sus	&	15.164	&	181.1	&		&	antiphase	\\
ZTF J201036.73+365925.3	&	20 10 36.74	&	+36 59 25.38	&	ZTF\_per	&	16.069	&	50.0	&		&	amp $r>g$	\\
ZTF J210819.95+482948.2	&	21 08 19.95	&	+48 29 48.41	&	ZTF\_per	&	15.836	&	64.5	&		&	amp $r>g$ \\
ZTF J211053.69+432024.2	&	21 10 53.69	&	+43 20 24.27	&	ZTF\_per	&	17.294	&	194.9	&		&	antiphase	\\
ZTF J212540.46+594620.2	&	21 25 40.48	&	+59 46 20.26	&	ZTF\_per	&	14.217	&	64.1	&		&	antiphase	\\
ZTF J212623.27+551353.6	&	21 26 23.28	&	+55 13 53.62	&	ZTF\_per	&	14.361	&	83.8	&		&	amp $r>g$	\\
ZTF J213025.59+475535.3	&	21 30 25.60	&	+47 55 35.27	&	ZTF\_per	&	14.627	&	110.5	&	kA0hB9 V SiSrCrEu	&	amp $r>g$	\\
ZTF J213425.23+504604.4	&	21 34 25.24	&	+50 46 04.46	&	ZTF\_per	&	15.117	&	140.8	&		&	antiphase	\\
ZTF J213435.94+595813.4	&	21 34 35.97	&	+59 58 13.51	&	ZTF\_per	&	13.850	&	89.1	&		&	antiphase	\\
ZTF J214050.19+521625.7	&	21 40 50.21	&	+52 16 25.67	&	ZTF\_per	&	15.843	&	117.5	&		&	antiphase	\\
ZTF J214501.59+643441.9	&	21 45 01.60	&	+64 34 41.96	&	ZTF\_per	&	15.020	&	61.9	&		&	antiphase	\\
ZTF J215742.24+492955.8	&	21 57 42.26	&	+49 29 55.86	&	ZTF\_per	&	13.213	&	115.4	&		&	antiphase	\\
ZTF J215751.18+562522.6	&	21 57 51.19	&	+56 25 22.56	&	ZTF\_per	&	15.153	&	177.9	&		&	amp $r\sim g$	\\
ZTF J220344.78+460557.5	&	22 03 44.79	&	+46 05 57.60	&	ZTF\_per	&	15.377	&	69.9	&		&	antiphase	\\
ZTF J221316.28+471544.8	&	22 13 16.30	&	+47 15 44.83	&	ZTF\_per	&	15.323	&	143.2	&		&	amp $r\sim g$	\\
ZTF J221537.26+513506.5	&	22 15 37.27	&	+51 35 06.59	&	ZTF\_per	&	12.849	&	183.3	&		&	amp $r>g$	\\
ZTF J223955.81+582547.7	&	22 39 55.82	&	+58 25 47.73	&	ZTF\_per	&	14.033	&	230.1	&		&	antiphase	\\
ZTF J224236.86+521540.8	&	22 42 36.86	&	+52 15 40.83	&	ZTF\_per	&	12.647	&	65.5	&		&	amp $r>g$	\\
ZTF J224614.86+540241.8	&	22 46 14.87	&	+54 02 41.84	&	ZTF\_per	&	13.399	&	59.4	&		&	LC shape	\\
\hline
\end{tabular} 
\end{adjustbox}
\setcounter{table}{1}
\end{table*}
\begin{table*}
\caption{Basic parameters for the newly identified sample of ssrAp stars and candidates (continued).}
\label{table_results2}
\begin{adjustbox}{max width=0.99\textwidth}
\begin{tabular}{lllllcll}
\hline
\hline
ID & RA (J2000) & Dec (J2000) & Source & $G$ mag & $P$ & Spectral type & Remark \\
\hline
ZTF J225959.83+610427.5	&	22 59 59.85	&	+61 04 27.47	&	ZTF\_per	&	15.878	&	110.4	& \hspace{90pt}	&	antiphase	\\
ZTF J230106.58+591736.7	&	23 01 06.59	&	+59 17 36.69	&	ZTF\_per	&	16.917	&	63.7	&		&	antiphase	\\
ZTF J230345.17+652605.3	&	23 03 45.18	&	+65 26 05.32	&	ZTF\_per	&	15.052	&	60.1	&		&	antiphase	\\
ZTF J230923.96+645158.7	&	23 09 23.97	&	+64 51 58.70	&	ZTF\_per	&	15.786	&	80.2	&		&	amp $r>g$	\\
ZTF J232607.29+611952.2	&	23 26 07.30	&	+61 19 52.24	&	ZTF\_per	&	15.665	&	141.0	&		&	antiphase	\\
ZTF J232932.44+612137.9	&	23 29 32.45	&	+61 21 37.87	&	ZTF\_per	&	16.036	&	95.9	&		&	antiphase	\\
ZTF J235136.41+490754.2	&	23 51 36.42	&	+49 07 54.32	&	ZTF\_per	&	14.192	&	128.9	&		&	antiphase	\\
ZTF J235247.14+593239.3	&	23 52 47.15	&	+59 32 39.27	&	ZTF\_per	&	16.147	&	68.7	&		&	amp $r>g$	\\
\hline
\end{tabular} 
\end{adjustbox}
\tablefoot{The columns denote: (1) Identifier; (2) Right ascension (J2000; Gaia DR3); (3) Declination (J2000; Gaia DR3); (4) Source; (5) $G$ mag (Gaia DR3); (6) Photometric variability period in days, as derived in this study; (7) Spectral type, where available, as derived in this study; (8) Remark.}
\end{table*}

\clearpage

\section{Phase plots}

\begin{figure*}
    \includegraphics[width=0.45\textwidth]{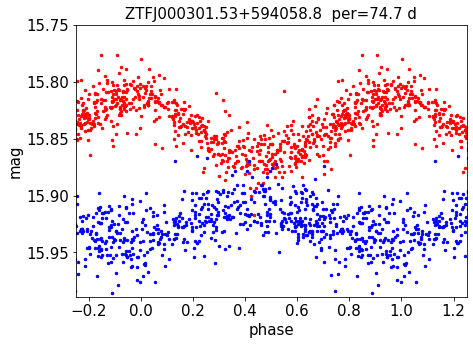}
    \includegraphics[width=0.45\textwidth]{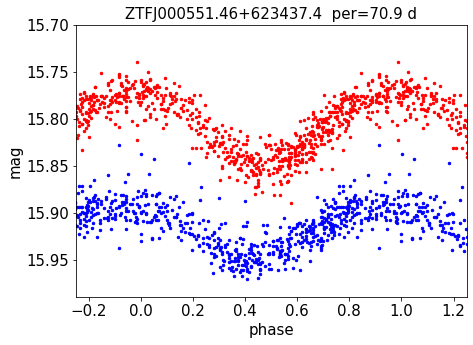} \\
    \includegraphics[width=0.45\textwidth]{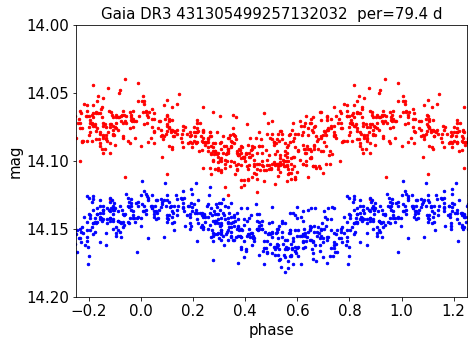}
    \includegraphics[width=0.45\textwidth]{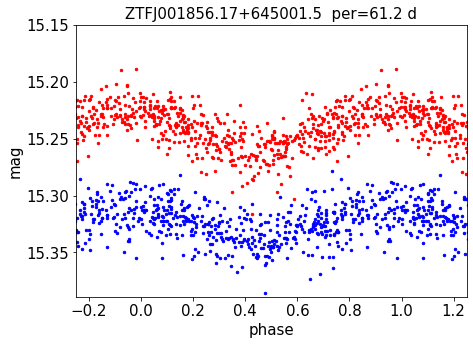} \\
    \includegraphics[width=0.45\textwidth]{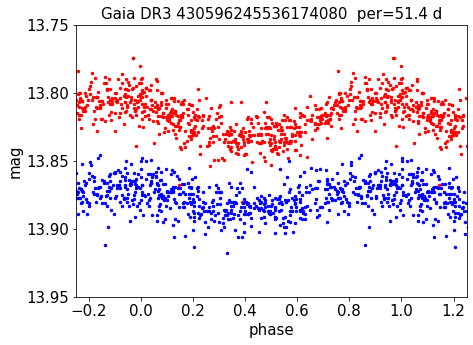}
    \includegraphics[width=0.45\textwidth]{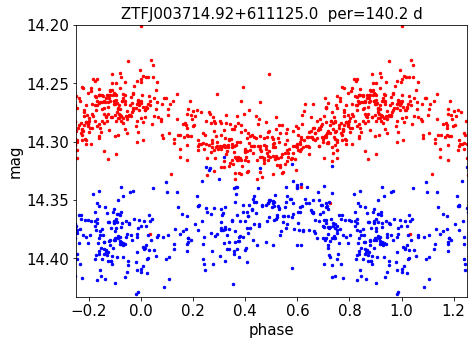} \\
    \includegraphics[width=0.45\textwidth]{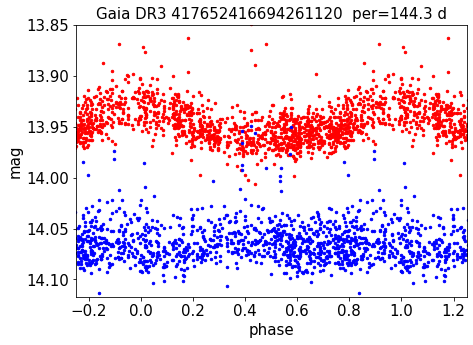}
    \includegraphics[width=0.45\textwidth]{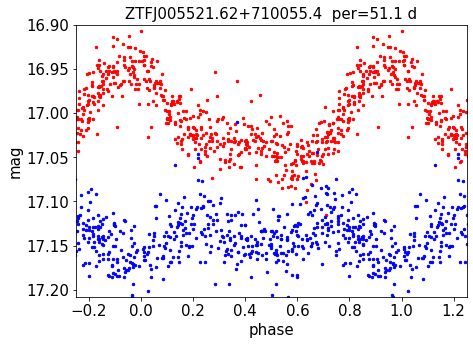} \
      \caption{Light curves of our target stars, phased with the periods listed in Table \ref{table_results1}. Red and blue symbols correspond to ZTF $r$ and $g$ band data, respectively. For clarity, $g$ band data have been shifted by varying amounts in the phase plots.}
         \label{fig:phaseplots1}
\end{figure*}
\setcounter{figure}{0}
\begin{figure*}
    \includegraphics[width=0.45\textwidth]{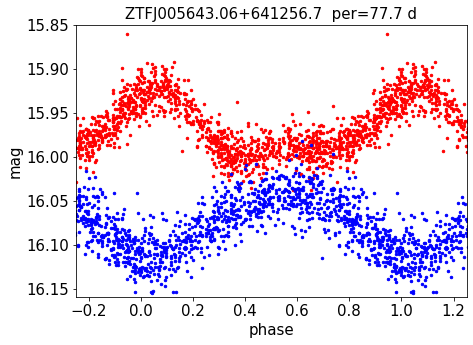}
    \includegraphics[width=0.45\textwidth]{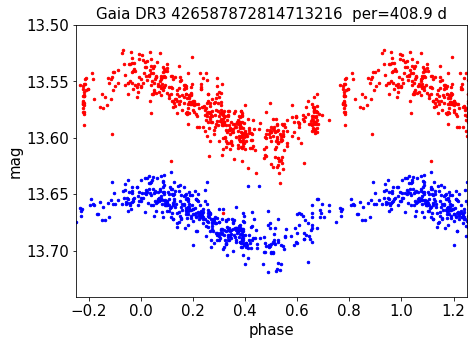} \\
    \includegraphics[width=0.45\textwidth]{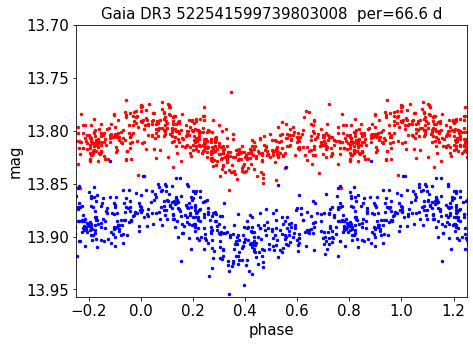}
    \includegraphics[width=0.45\textwidth]{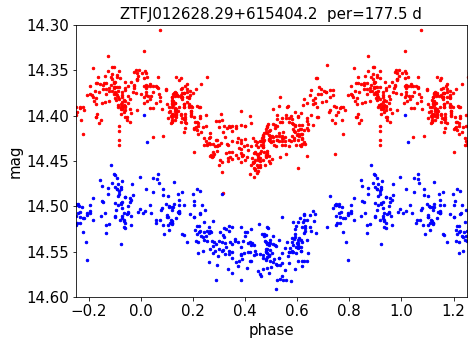} \\
    \includegraphics[width=0.45\textwidth]{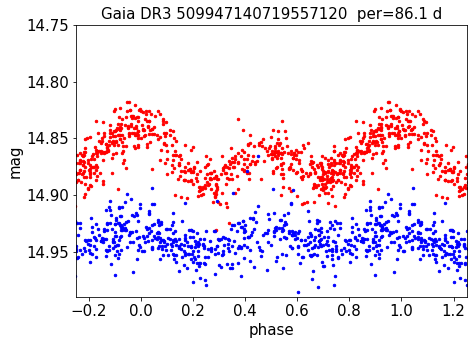}
    \includegraphics[width=0.45\textwidth]{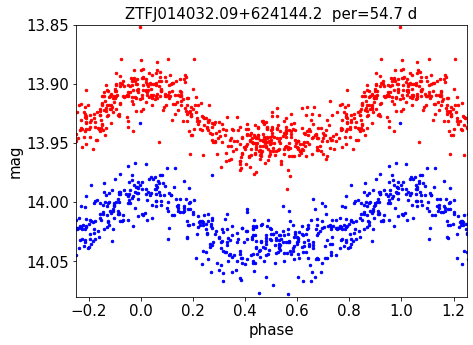} \\
    \includegraphics[width=0.45\textwidth]{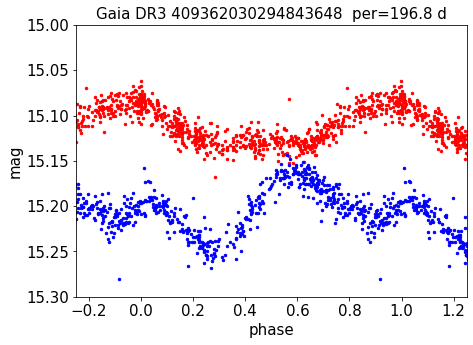}
    \includegraphics[width=0.45\textwidth]{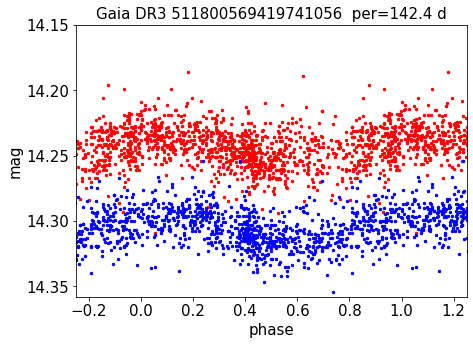} \
      \caption{Continued.}
         \label{fig:phaseplots2}
\end{figure*}
\setcounter{figure}{0}
\begin{figure*}
    \includegraphics[width=0.45\textwidth]{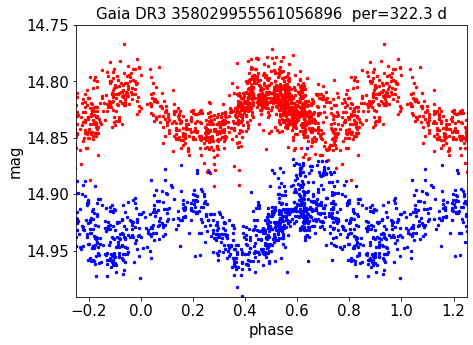}
    \includegraphics[width=0.45\textwidth]{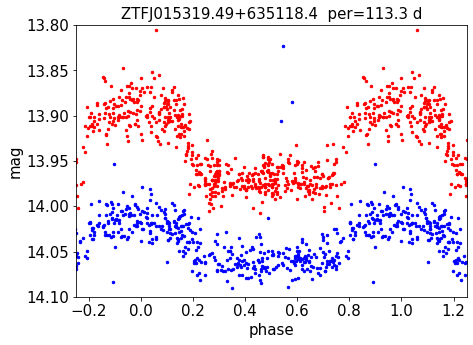} \\
    \includegraphics[width=0.45\textwidth]{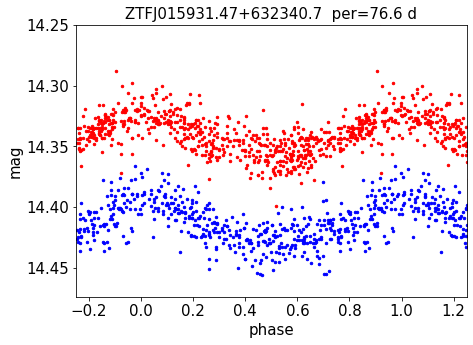}
    \includegraphics[width=0.45\textwidth]{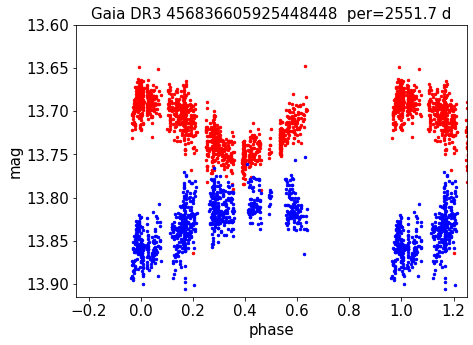} \\
    \includegraphics[width=0.45\textwidth]{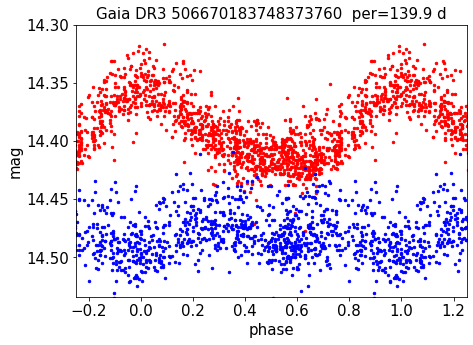}
    \includegraphics[width=0.45\textwidth]{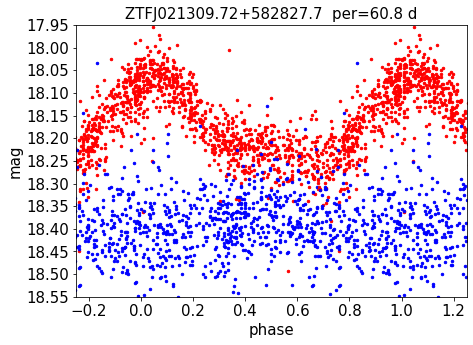} \\
    \includegraphics[width=0.45\textwidth]{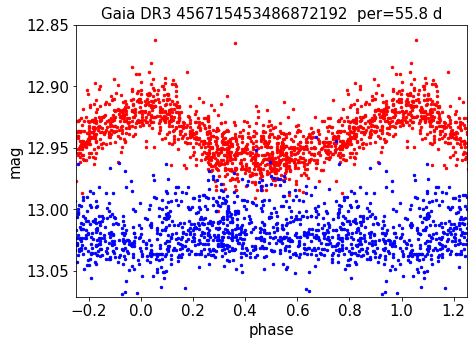}
    \includegraphics[width=0.45\textwidth]{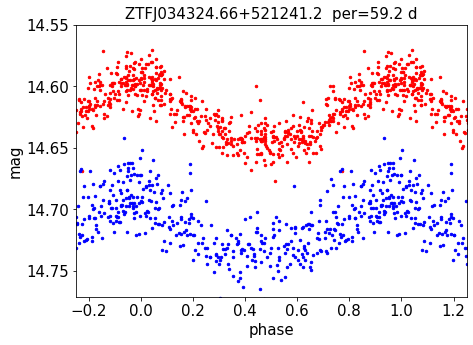} \
      \caption{Continued.}
         \label{fig:phaseplots3}
\end{figure*}
\setcounter{figure}{0}
\begin{figure*}
    \includegraphics[width=0.45\textwidth]{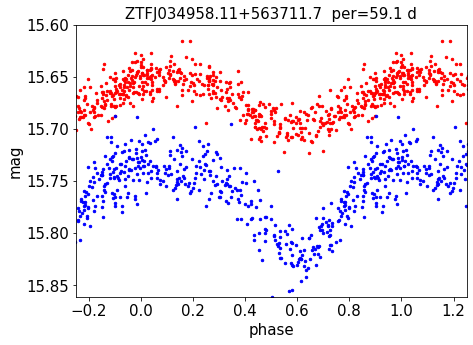}
    \includegraphics[width=0.45\textwidth]{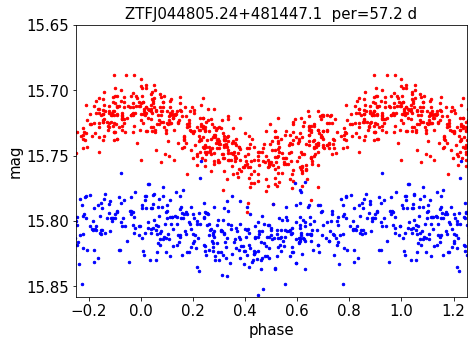} \\
    \includegraphics[width=0.45\textwidth]{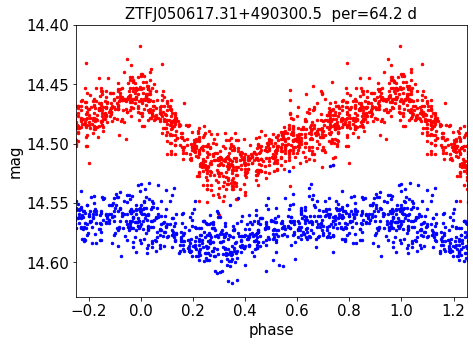}
    \includegraphics[width=0.45\textwidth]{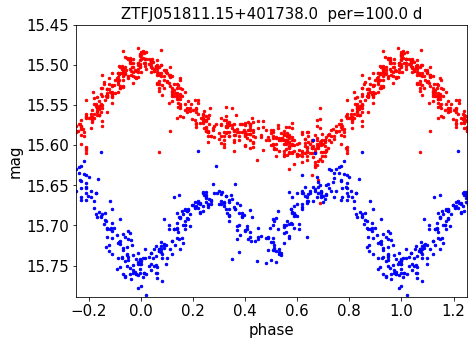} \\
    \includegraphics[width=0.45\textwidth]{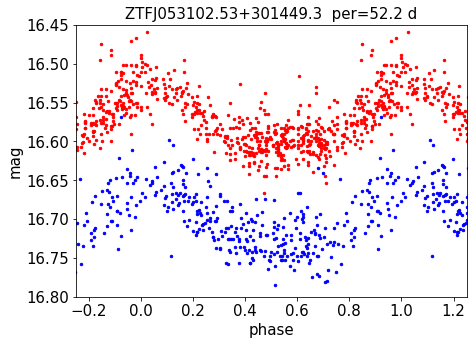}
    \includegraphics[width=0.45\textwidth]{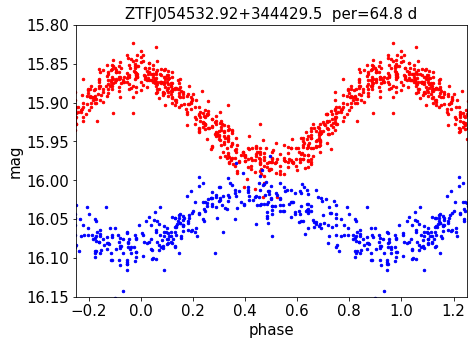} \\
    \includegraphics[width=0.45\textwidth]{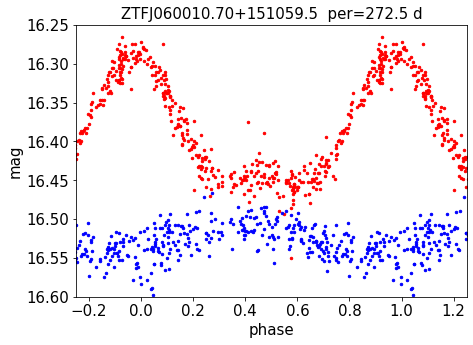}
    \includegraphics[width=0.45\textwidth]{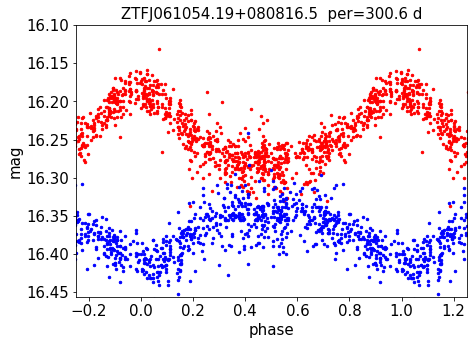} \
      \caption{Continued.}
         \label{fig:phaseplots4}
\end{figure*}\setcounter{figure}{0}
\begin{figure*}
    \includegraphics[width=0.45\textwidth]{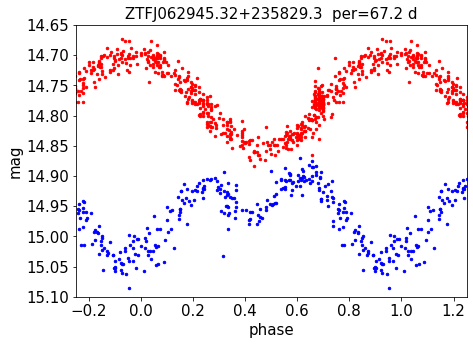}
    \includegraphics[width=0.45\textwidth]{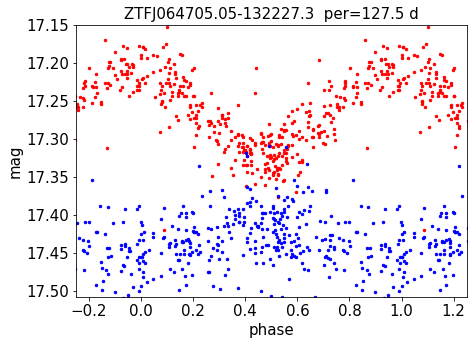} \\
    \includegraphics[width=0.45\textwidth]{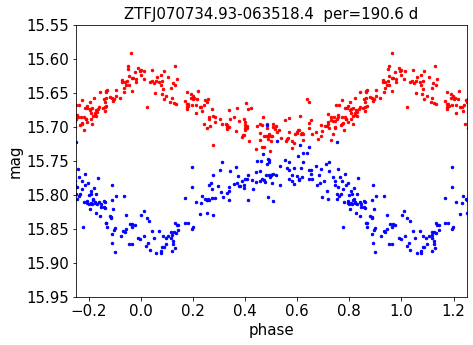}
    \includegraphics[width=0.45\textwidth]{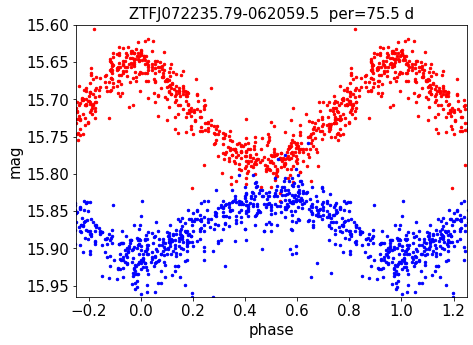} \\
    \includegraphics[width=0.45\textwidth]{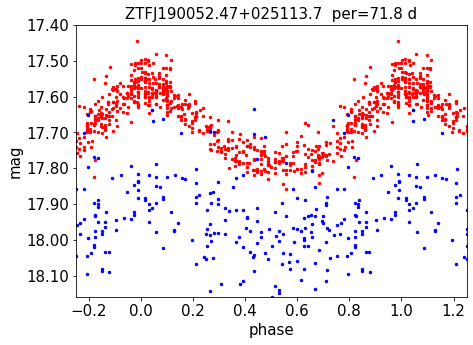}
    \includegraphics[width=0.45\textwidth]{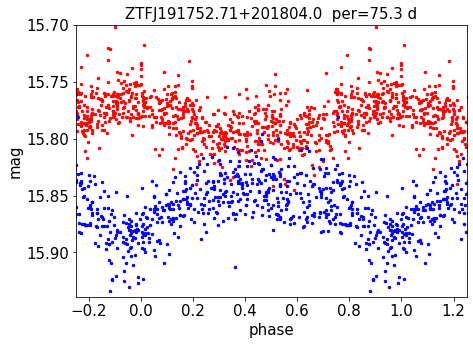} \\
    \includegraphics[width=0.45\textwidth]{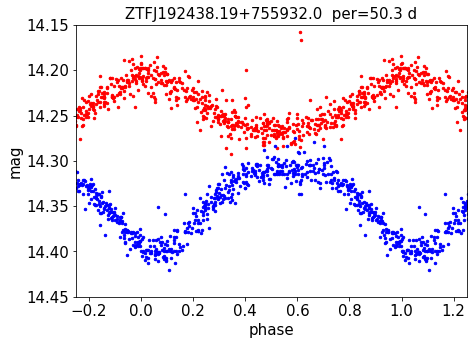}
    \includegraphics[width=0.45\textwidth]{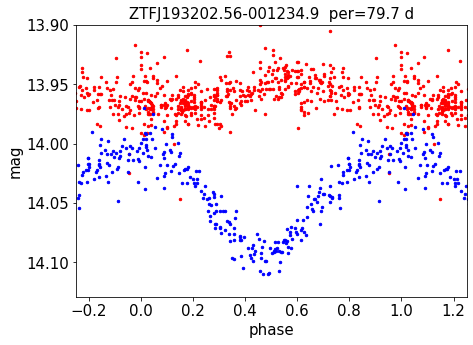} \
      \caption{Continued.}
         \label{fig:phaseplots5}
\end{figure*}
\setcounter{figure}{0}
\begin{figure*}
    \includegraphics[width=0.45\textwidth]{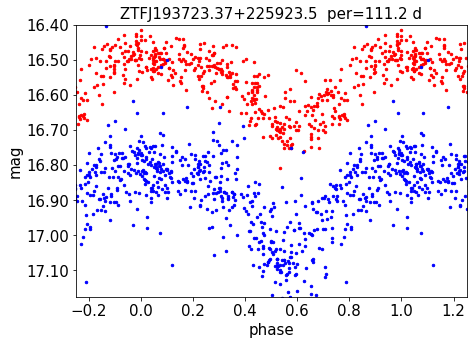}
    \includegraphics[width=0.45\textwidth]{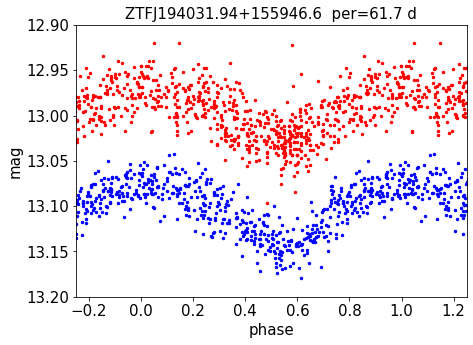} \\
    \includegraphics[width=0.45\textwidth]{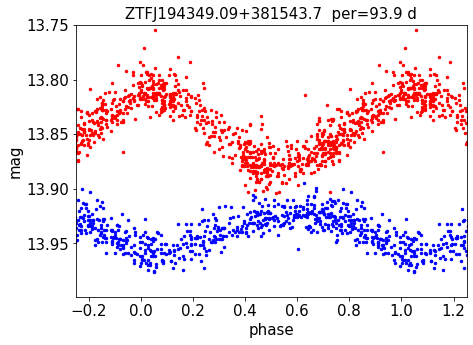}
    \includegraphics[width=0.45\textwidth]{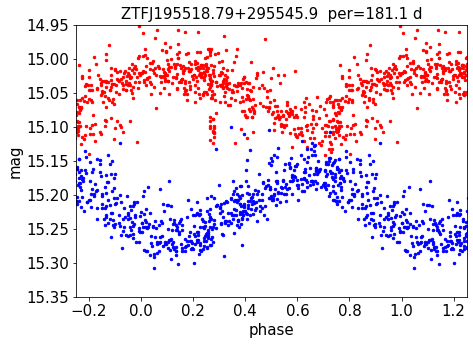} \\
    \includegraphics[width=0.45\textwidth]{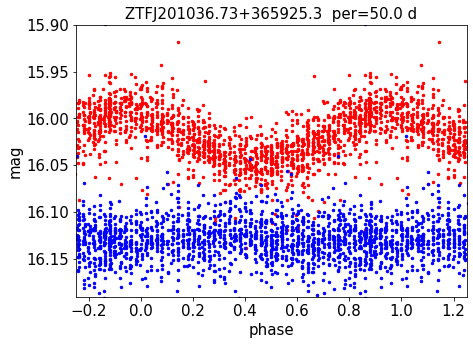}
    \includegraphics[width=0.45\textwidth]{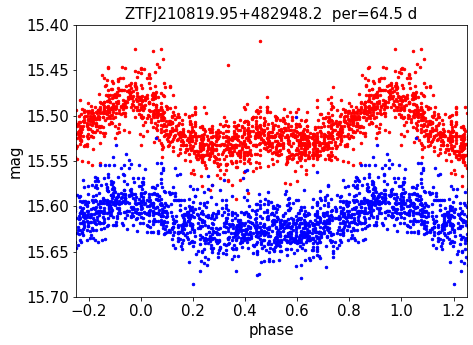} \\
    \includegraphics[width=0.45\textwidth]{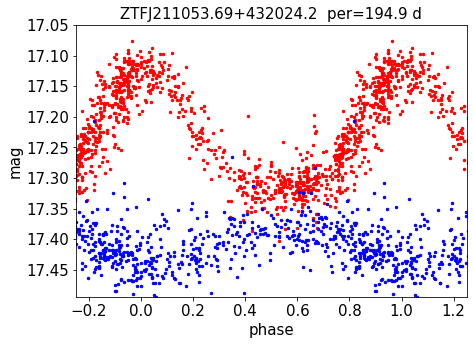}
    \includegraphics[width=0.45\textwidth]{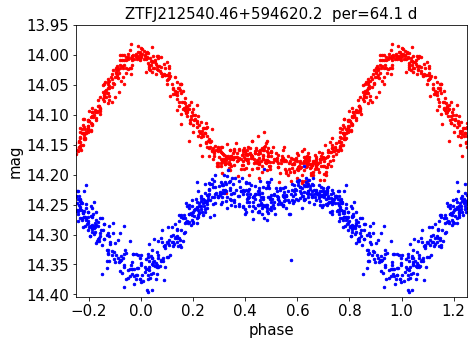} \
      \caption{Continued.}
         \label{fig:phaseplots6}
\end{figure*}
\setcounter{figure}{0}
\begin{figure*}
    \includegraphics[width=0.45\textwidth]{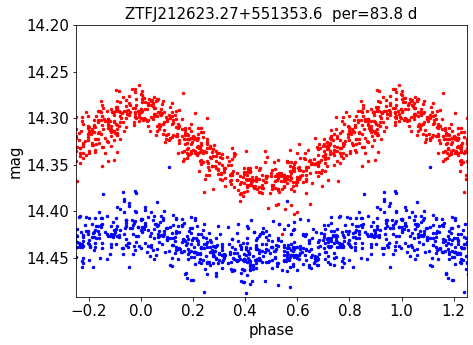}
    \includegraphics[width=0.45\textwidth]{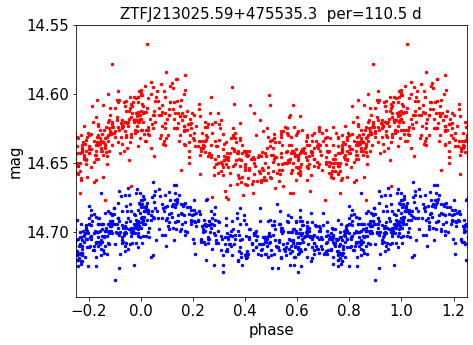} \\
    \includegraphics[width=0.45\textwidth]{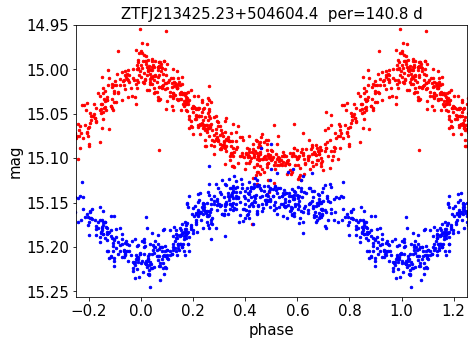}
    \includegraphics[width=0.45\textwidth]{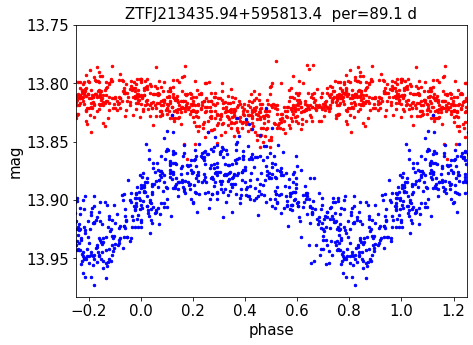} \\
    \includegraphics[width=0.45\textwidth]{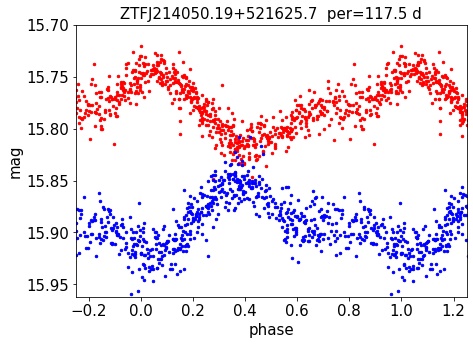}
    \includegraphics[width=0.45\textwidth]{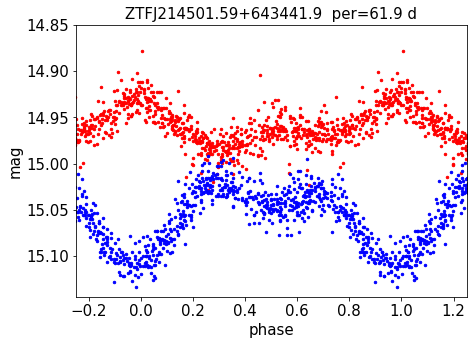} \\
    \includegraphics[width=0.45\textwidth]{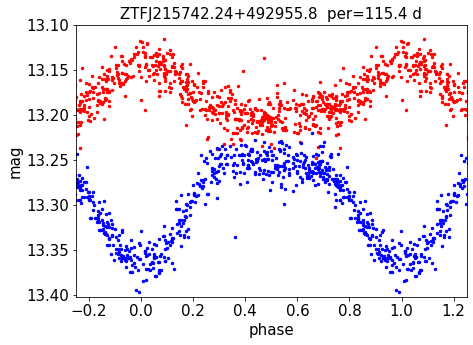}
    \includegraphics[width=0.45\textwidth]{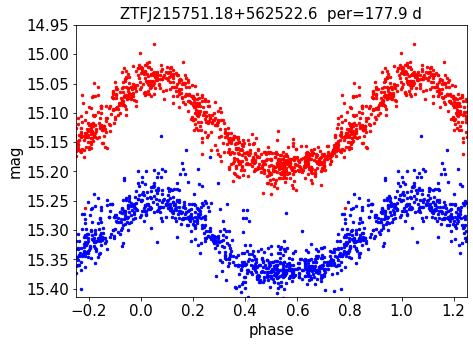} \
      \caption{Continued.}
         \label{fig:phaseplots7}
\end{figure*}
\setcounter{figure}{0}
\begin{figure*}
    \includegraphics[width=0.45\textwidth]{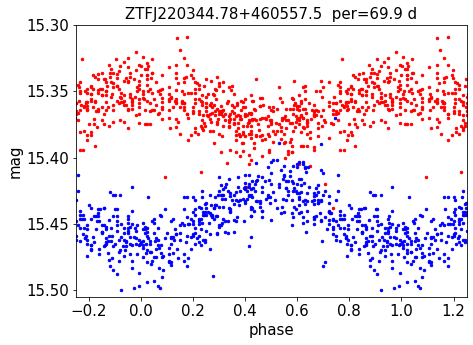}
    \includegraphics[width=0.45\textwidth]{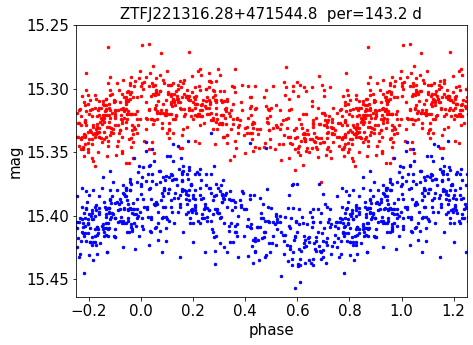} \\
    \includegraphics[width=0.45\textwidth]{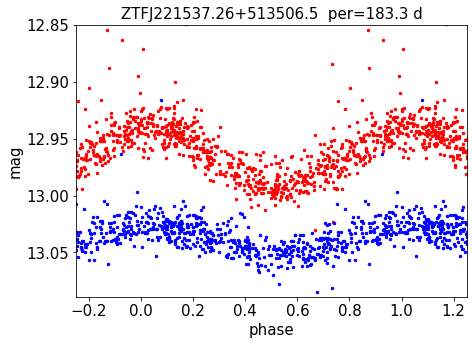}
    \includegraphics[width=0.45\textwidth]{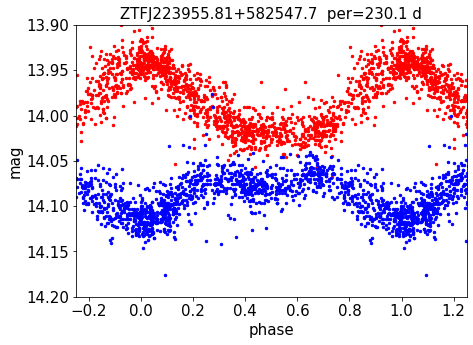} \\
    \includegraphics[width=0.45\textwidth]{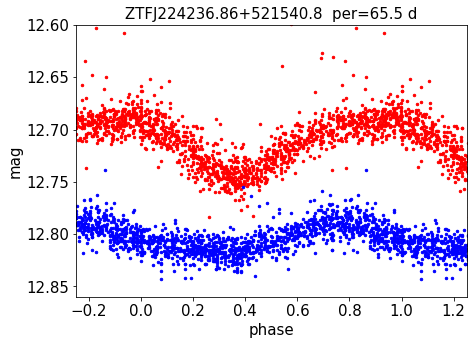}
    \includegraphics[width=0.45\textwidth]{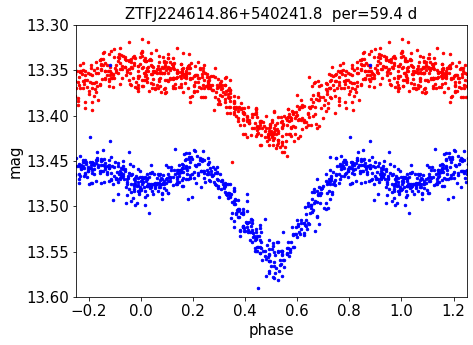} \\
    \includegraphics[width=0.45\textwidth]{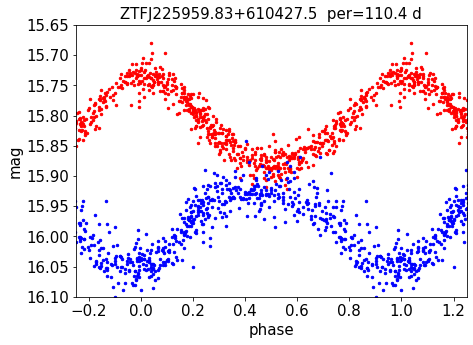}
    \includegraphics[width=0.45\textwidth]{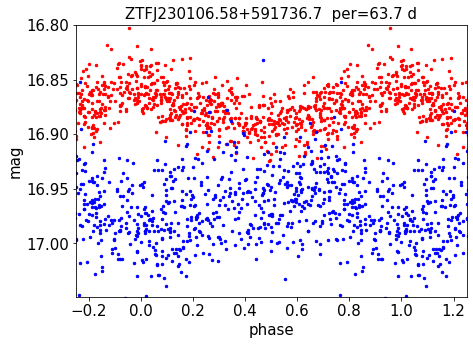} \
      \caption{Continued.}
         \label{fig:phaseplots8}
\end{figure*}
\setcounter{figure}{0}
\begin{figure*}
    \includegraphics[width=0.45\textwidth]{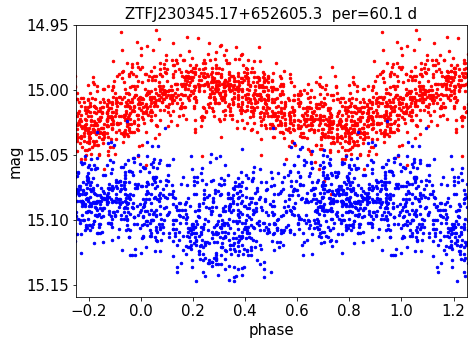}
    \includegraphics[width=0.45\textwidth]{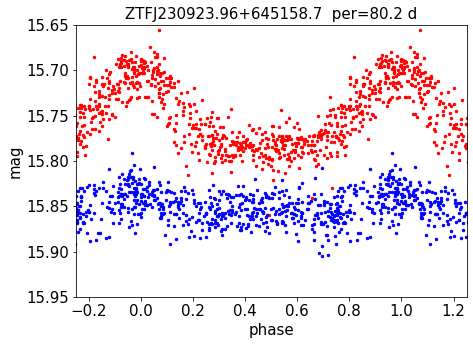} \\
    \includegraphics[width=0.45\textwidth]{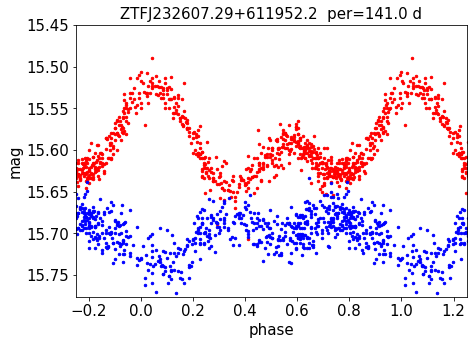}
    \includegraphics[width=0.45\textwidth]{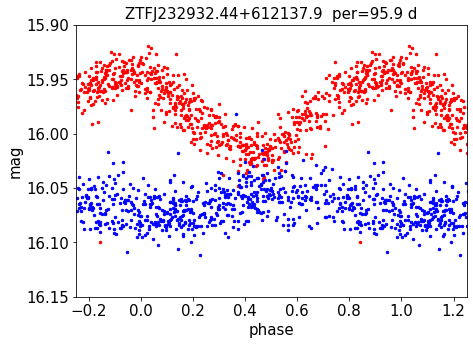} \\
    \includegraphics[width=0.45\textwidth]{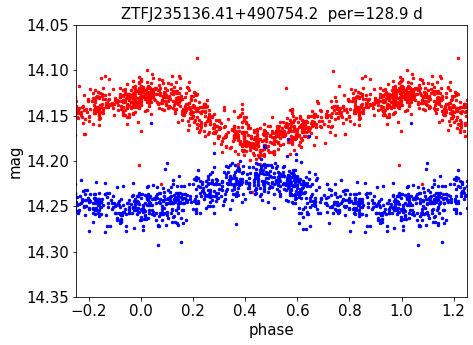}
    \includegraphics[width=0.45\textwidth]{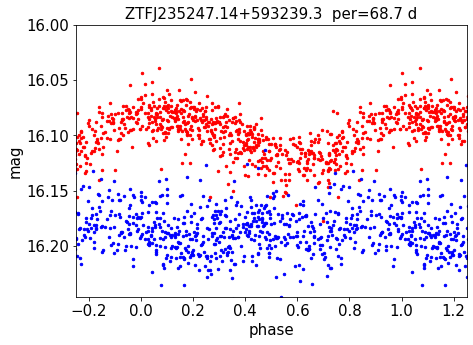} \\
      \caption{Continued.}
         \label{fig:phaseplots9}
\end{figure*}

\end{appendix}

\end{document}